\documentclass[12pt,letterpaper]{article}

\usepackage{amssymb, amsmath, amsthm}
\usepackage{bm}
\usepackage{booktabs}
\usepackage{graphicx}
\usepackage{placeins, flafter}
\usepackage{natbib}
\usepackage{setspace}
\usepackage{float}
\usepackage{url}
\newtheorem{proposition}{Proposition}
\newtheorem{theorem}{Theorem}

\newtheorem{assumption}{Assumption}

\newcommand{\di}{\text{d}}
\newcommand{\bh}{\mathbf{h}}
\newcommand{\by}{\mathbf{y}}

\newcommand{\bu}{\mathbf{u}}
\newcommand{\bw}{\mathbf{w}}

\newcommand{\bA}{\mathbf{A}}
\newcommand{\bLam}{\bm{\Lambda}}
\newcommand{\bG}{\mathbf{G}}
\newcommand{\bI}{\mathbf{I}}
\newcommand{\bK}{\mathbf{K}}
\newcommand{\bD}{\mathbf{D}}
\newcommand{\ba}{\mathbf{a}}
\newcommand{\be}{\mathbf{e}}
\newcommand{\bone}{\mathbf{1}}
\newcommand{\distn}[1]{\mathcal{#1}}
\newcommand{\Em}{\mathbb E}

\newcommand{\tbf}{\text{BF}}
\newcommand{\gvn}{\,|\,}
\newcommand{\e}{\text{e}}
\newcommand{\Var}{\text{Var}}
\newcommand{\Cov}{\text{Cov}}
\newcommand{\diag}{\text{diag}}

\newcommand{\cum}{\text{cum}}
\newcommand{\matlab}{\mathrm{M}\mathrm{{\scriptstyle ATLAB}}}
\renewcommand{\epsilon}{\varepsilon}
\renewcommand{\hat}{\widehat}
\renewcommand{\tilde}{\widetilde}
\renewcommand{\leq}{\leqslant}
\renewcommand{\geq}{\geqslant}

\begin{document}

\title{Bayesian Model Comparison for Nonlinear State Space Models without Posterior Sampling}

\author{Joshua C.C. Chan \\
 {\small Purdue University}
 \and
 Wei Zhang\thanks{Wei Zhang gratefully acknowledges financial support through her postdoctoral fellowship at the Global Macro Lab, Johns Hopkins University.} \\
 {\small Johns Hopkins University}
}

\date{November 2026}

\maketitle

\doublespacing

\begin{abstract}

\noindent Marginal likelihoods are central to Bayesian comparison of nonlinear state space models, but commonly used estimators require posterior simulation for each model, with computational costs that grow with the sample size $T$. We propose a deterministic estimator that requires no posterior simulation. A grid filter integrates out the latent states, and a Laplace approximation with a second-order correction integrates over the static parameters. Under derivative-growth and posterior-concentration conditions, the corrected Laplace approximation has error $O(T^{-2})$, and the size of the correction indicates the error that remains. In stochastic volatility experiments, the estimator is more accurate than Chib's method, the Gelfand--Dey estimator and importance sampling with a fitted proposal at equal computing time for $T\geq 2{,}000$. Applications to two exchange rates and to $T=19{,}266$ five-minute stock returns provide the strongest support for heavy-tailed observations and a transitory volatility factor.

\bigskip

\noindent \textbf{JEL classification:} C11, C15, C52, C58

\bigskip

\noindent \textbf{Keywords:} marginal likelihood, Bayesian model comparison, Laplace approximation, nonlinear state space models, stochastic volatility, high-frequency data

\end{abstract}

\thispagestyle{empty}

\newpage

\section{Introduction} \label{s:intro}

Nonlinear state space models are widely used to study time-varying volatility, counts, durations, and other latent dynamics in macroeconomics and finance. High-frequency financial data can supply tens of thousands of observations on a single asset. Models of volatility spillovers, heavy tails, jumps and multiple volatility factors require a few latent processes and non-Gaussian observations. Empirical work then has to compare many competing specifications on long series. These models have two sets of unknowns: a latent state path, whose dimension grows with the sample size $T$, and a small vector of static parameters $\bm{\theta}$. Bayesian model comparison is based on the marginal likelihood, which integrates over both. Integrating out the states gives the integrated likelihood $Z_T(\bm{\theta}) = p(\by \gvn \bm{\theta})$, and integrating this against the prior density $p(\bm{\theta})$ gives the marginal likelihood, $m_T = \int Z_T(\bm{\theta})p(\bm{\theta})\,\di\bm{\theta}$. 

Computing this nested integral accurately is challenging. In nonlinear and non-Gaussian models, $Z_T(\bm{\theta})$ is generally unavailable analytically and must be evaluated by filtering or simulation over the full sample. Then computing the marginal likelihood adds an outer integration over the static parameters, which is costly to evaluate. Simulation-based methods require either repeated evaluations of $Z_T(\bm{\theta})$ across parameter values or posterior simulation over the parameters and latent states, with a sampler written for each model. Each sweep and each likelihood evaluation takes longer as $T$ grows, so the computing time for a given accuracy grows with $T$.

We propose a deterministic estimator that requires no posterior sampler and computes the marginal likelihood in two steps. First, a numerical-integration filter on a grid integrates out the states, exploiting the Markov structure of the state space model. Although the number of latent states grows with the sample size $T$, the number of static parameters remains fixed. Integrating out the states leaves a fixed, low-dimensional integral over the static parameters, which the second step approximates by a Laplace expansion around the posterior mode.  Under regularity conditions, its accuracy improves as the posterior of the static parameters concentrates with increasing $T$. In our implementation, mode finding and derivative calculations require a few hundred to several thousand likelihood evaluations, many of which can run in parallel batches. Therefore, the estimator is particularly attractive for long samples, where simulation-based integration is most costly.

At moderate sample sizes, however, skewness and other departures from posterior normality can limit the accuracy of the ordinary Laplace approximation. We therefore add a second-order correction based on the third and fourth derivatives of the log integrated likelihood and the log prior at the posterior mode. This correction captures the leading effects of posterior nonnormality while retaining the computational advantages of a local approximation. It also acts as a diagnostic, because the error that remains after the correction is of the order of its square: we show in the Monte Carlo experiments that the absolute error of the corrected log marginal likelihood is between 0.5 and 7 times the squared correction in all 300 data sets. A small correction indicates an accurate estimate, and a correction of order one can serve as an indicator that the expansion is not reliable along some direction of the parameter space. In that case, the mode and curvature already computed provide the proposal for importance sampling over the static parameters, which replaces the expansion.

Our theoretical analysis characterizes the correction and its error in terms of the integrated likelihood. We express the third and fourth derivatives of its logarithm through expectations and joint cumulants under the smoothing distribution, which links the correction to the latent-state structure. Under explicit derivative-growth and posterior-concentration conditions, our main theorem shows that the correction reduces the approximation error in the log marginal likelihood from order $T^{-1}$ to order $T^{-2}$. We separate this approximation error from the numerical errors in the mode, curvature and derivatives. The theory applies to a fixed, low-dimensional parameter vector with a posterior concentrated around a unique interior mode. The states are integrated out by a numerical-integration filter on a grid \citep{Kitagawa1987, FridmanHarris1998}. For a scalar or bivariate state it gives the integrated likelihood to machine precision, so the Laplace expansion is the only source of error of the estimator.

In Monte Carlo experiments on stochastic volatility and three related models, we measure the approximation error against quadrature benchmarks accurate to $10^{-4}$. The correction removes 64\% to 88\% of the Laplace error at $T = 500$ and 94\% to 99\% at $T = 8{,}000$, leaving log errors below 0.08 and 0.003. At equal computing time, the proposed estimator is more accurate than Chib's method, the Gelfand--Dey estimator and importance sampling with the cross-entropy method from $T = 2{,}000$ at every persistence level. At $T = 8{,}000$ its error is 9 to 400 times smaller than the standard error of the cross-entropy method, the best of the three. The errors of Chib's method and the Gelfand--Dey estimator rise steeply with $T$, because each MCMC sweep takes longer. On data generated from three bivariate stochastic volatility models, the proposed estimator selects the true model in 149 of 150 series; the exception is a near-tie on data from the simplest model, which the other two include as a special case.

We then present two applications. The first compares three bivariate stochastic volatility specifications drawn from \citet{YuMeyer2006} using $T = 4{,}000$ daily AUD/USD and NZD/USD returns. Starting from a constant-correlation baseline, the other two models separately introduce volatility spillovers and heavy-tailed observations. \citet{YuMeyer2006} compare their specifications by the deviance information criterion on weekly exchange-rate data, and to our knowledge marginal likelihoods have not previously been reported for them. In our sample, the heavy-tailed model receives the strongest support, with a log Bayes factor of 110 relative to the baseline, against 98 for volatility spillovers. The second-order term is below 0.2 for every model, so the remaining approximation error is far smaller than these differences.

The second application compares seven stochastic volatility models for $T = 19{,}266$ five-minute returns on the S\&P 500 ETF over 2025: the standard stochastic volatility model and its extensions with heavy tails, with jumps and with a second volatility factor. The two-factor model is the most favored by the data. The two extensions that change the dynamics of the volatility---a transitory factor and a jump in the volatility---have log Bayes factors of 69 to 73 relative to the baseline, against 12 to 25 for heavy tails and return jumps. The second-order term is at most 0.31, so the remaining approximation error is far smaller than these differences.

A key advantage of the proposed approach is that comparing many models on many samples is feasible. We therefore fit all seven models at the end of each month, 77 fits in all, to see whether the ranking is stable over time. The ranking does change through the year: the model with jumps in volatility is favored most strongly after the April and October tariff episodes, and the two-factor model from June to August. The second-order term identifies three of the 77 fits at which the expansion is unreliable, each time for the variance of rare jumps. Because the mode and curvature at those fits are already available, the proposal for importance sampling over the static parameters can be constructed at no additional cost, and the resulting estimator is, to our knowledge, new to nonlinear state space model comparison.

The paper contributes to two strands of literature. The first is simulation-based marginal likelihood estimation, including Chib's method \citep{chib95, CJ01}, bridge sampling \citep{MengWong1996}, importance sampling \citep{Geweke1989, CE15, CYZ26} and sequential Monte Carlo \citep{Creal2012, ChopinJacobPapaspiliopoulos2013}. All of these methods simulate over the parameters, and most require posterior draws from MCMC. Our contribution is to approximate the integral over the static parameters directly by a Laplace approximation with a second-order correction.

The second is the literature on Laplace approximations and posterior expansions. The second-order correction is the next term of Laplace's method \citep{TierneyKadane1986, KassTierneyKadane1990, ShunMcCullagh1995}; our contribution is its formulation and error analysis for nonlinear state space models after the states have been integrated out, with the derivatives expressed as smoothing cumulants. This differs from integrated nested Laplace approximation \citep{RueMartinoChopin2009}, which approximates the latent-field integral first, and from the higher-order conditional Laplace approach of \citet*{RuliSartoriVentura2016}. In score-driven filtering, \citet{HansenTong2026} show that the implicit update of \citet*{LangeVanOsVanDijk2026} is a local approximation to the Bayesian update of a time-varying parameter; our expansion concerns the integral over the static parameters.

The rest of the paper is organized as follows. Section~\ref{s:estimator} describes the proposed approach. Section~\ref{s:theory} states the theoretical results. Section~\ref{s:MC} measures the accuracy of the proposed estimator on generated data and how often it selects the true model. Sections~\ref{s:app} and~\ref{s:app2} present the applications, and Section~\ref{s:conclusion} concludes.

\section{Marginal Likelihood Approximation} \label{s:estimator}

The proposed approach combines a grid filter for the integrated likelihood with a Laplace approximation to the integral over the static parameters. We first formulate this integral in unconstrained coordinates. Then we derive the Laplace approximation and its second-order correction in standardized coordinates, and describe the grid filter that evaluates the integrated likelihood at the mode. Finally, we explain how the mode and the derivatives are computed, how the approximation error enters the estimator, and when importance sampling over the static parameters takes the place of the expansion.

Let $\by = (\by_1', \ldots, \by_T')'$ denote the observations and $\bm{\alpha} = (\bm{\alpha}_1', \ldots, \bm{\alpha}_T')'$ a latent state path. Conditional on a static parameter vector $\bm{\theta}$, the states form a Markov chain with transition density $p(\bm{\alpha}_t \gvn \bm{\alpha}_{t-1}, \bm{\theta})$ and initial density $p(\bm{\alpha}_1 \gvn \bm{\theta})$. The observations are conditionally independent given the states, with density $p(\by_t \gvn \bm{\alpha}_t, \bm{\theta})$. The integrated likelihood and the marginal likelihood are, respectively, 
\[
	Z_T(\bm{\theta}) = \int p(\by \gvn \bm{\alpha}, \bm{\theta})\,p(\bm{\alpha} \gvn \bm{\theta})\,\di \bm{\alpha},
	\qquad
	m_T = \int Z_T(\bm{\theta})\,p(\bm{\theta})\,\di \bm{\theta},
\]
where $p(\bm{\theta})$ is a proper prior. Model comparison needs to use the integrated likelihood $Z_T$, because criteria computed from the likelihood conditional on the states are unreliable \citep[see, e.g.,][]{CG15, LiYuZeng2020}.

Our running example is the stochastic volatility (SV) model
\begin{equation} \label{eq:sv}
	y_t = \e^{h_t/2}\epsilon_t, \qquad h_t = \mu + \phi(h_{t-1} - \mu) + \sigma\eta_t, \qquad h_1 \sim \distn{N}\big(\mu, \sigma^2/(1 - \phi^2)\big),
\end{equation}
with $\epsilon_t$ and $\eta_t$ independent standard normal. The state is the scalar log-volatility $h_t$, and the static parameters are $\bm{\theta} = (\mu, \phi, \sigma^2)'$, where $\phi$ and $\sigma$ have the constrained supports $|\phi| < 1$ and $\sigma > 0$.

A Gaussian approximation, such as that underlying the Laplace method, does not respect these constrained supports. Therefore, we transform the static parameters to unconstrained coordinates to remove parameter boundaries and potentially reduce posterior asymmetry, thereby improving approximation accuracy. Specifically, let $\bm{\psi} = g(\bm{\theta}) \in \mathbb{R}^d$ be a smooth bijection, and let $p_{\bm{\psi}}(\bm{\psi})$ denote the transformed prior density. In these coordinates, define the log posterior kernel as
\[
	f(\bm{\psi}) = \log Z_T(\bm{\psi}) + \log p_{\bm{\psi}}(\bm{\psi}),
\]
where $Z_T(\bm{\psi})$ denotes the integrated likelihood evaluated at $\bm{\theta} = g^{-1}(\bm{\psi})$. The marginal likelihood can then be written as $m_T = \int_{\mathbb{R}^d} \e^{f(\bm{\psi})}\,\di \bm{\psi}.$ For the stochastic volatility model \eqref{eq:sv}, we use $\bm{\psi} = (\mu, \text{atanh}\,\phi, \log\sigma^2)'$, which maps the restricted supports of the persistence parameter $\phi \in (-1,1)$ and the variance $\sigma^2 > 0$ to the real line. Online Appendix~\ref{app:implementation} records all transformations and their Jacobian terms used in this paper.

\subsection{Laplace Approximation and the Second-Order Correction} \label{ss:laplace}

Let $\hat{\bm{\psi}}$ denote the posterior mode, the maximizer of the log posterior kernel $f$. Define $\bLam = -\nabla^2 f(\hat{\bm{\psi}})$ as the negative Hessian matrix at the mode, assumed positive definite, and let $\bA = \bLam^{-1/2}$ be its symmetric inverse square root. The ordinary Laplace approximation replaces the log posterior kernel $f$ by its quadratic expansion around the mode and integrates the resulting Gaussian kernel. This gives the approximation $m_T \approx m_1$, where
\begin{equation} \label{eq:m0}
	\begin{aligned}
		m_1
		&= \int_{\mathbb{R}^d}
		\exp\left\{
		f(\hat{\bm{\psi}})
		-\frac12(\bm{\psi}-\hat{\bm{\psi}})'
		\bLam(\bm{\psi}-\hat{\bm{\psi}})
		\right\}\,\di\bm{\psi} \\
		&= (2\pi)^{d/2}|\bLam|^{-1/2}\,\e^{f(\hat{\bm{\psi}})} 
		= (2\pi)^{d/2}|\bLam|^{-1/2}\,
		p_{\bm{\psi}}(\hat{\bm{\psi}})\,Z_T(\hat{\bm{\psi}}).
	\end{aligned}
\end{equation}
This approximation is exact when $f$ is globally quadratic. Otherwise, its error reflects the terms omitted from this
quadratic expansion. To isolate these terms, introduce the standardized coordinates
$\bu=\bA^{-1}(\bm{\psi}-\hat{\bm{\psi}})$, or equivalently
$\bm{\psi}=\hat{\bm{\psi}}+\bA\bu$. Define
\[
R(\bu)
=
\underbrace{f(\hat{\bm{\psi}}+\bA\bu)}_{\text{log posterior kernel}}
-
\underbrace{\left[f(\hat{\bm{\psi}})-\tfrac12\bu'\bu\right]}_{\text{quadratic approximation}}
\]
as the difference between the log posterior
kernel and its quadratic approximation in standardized coordinates. It can be shown that the ratio of the marginal likelihood to its ordinary Laplace approximation is
\begin{equation} \label{eq:ratio}
	\frac{m_T}{m_1} = \Em\big[\e^{R(\bG)}\big],
\end{equation}
where $\bG \sim \distn{N}(\bm{0}, \bI_d)$. The correction follows by expanding $R$ and integrating the resulting terms against the standard Gaussian density \citep[Chapter~6]{BarndorffNielsenCox1989}; \citet*{KassTierneyKadane1990} use the same expansion, in coordinates rescaled by the square root of the sample size, to establish the validity of posterior expansions. Because the linear term vanishes at the mode and the quadratic term has been removed, the Taylor expansion starts at order three:
\[
	R(\bu) = C_3(\bu) + C_4(\bu) + \cdots, \qquad C_k(\bu) = \frac{1}{k!}\,D^k f(\hat{\bm{\psi}})[\bA\bu, \ldots, \bA\bu],
\]
where $D^k f(\hat{\bm{\psi}})[\cdot]$ is the $k$th derivative of $f$ at $\hat{\bm{\psi}}$ as a multilinear form. The coefficients of $C_3$ and $C_4$ are the standardized derivative tensors
\[
	a_{ijk} = D^3 f(\hat{\bm{\psi}})[\bA\be_i, \bA\be_j, \bA\be_k],
	\qquad
	b_{ijkl} = D^4 f(\hat{\bm{\psi}})[\bA\be_i, \bA\be_j, \bA\be_k, \bA\be_l],
\]
where $\be_i$ is the $i$th column of $\bI_d$, so that $\bA\be_i$ is the $i$th column of $\bA$. Taking the expectation in \eqref{eq:ratio} term by term, the odd-order terms vanish by the symmetry of the Gaussian, and the leading correction to $m_T/m_1 = 1$ is $\Em[C_4(\bG)] + \tfrac12\Em[C_3(\bG)^2]$. In terms of the tensors, this second-order correction is
\begin{equation} \label{eq:delta2}
	\Delta_2 = \frac18 \sum_{i,j} b_{iijj} + \frac1{12}\sum_{i,j,k} a_{ijk}^2 + \frac18\sum_i\Big(\sum_j a_{ijj}\Big)^2,
\end{equation}
where the first sum is $\Em[C_4(\bG)]$ and the second and third sums together are $\tfrac12\Em[C_3(\bG)^2]$, the two contractions of the square of the cubic form under the Gaussian sixth moment. In the usual case in which $\bLam$ grows linearly in $T$, the standardized tensors satisfy $a_{ijk} = O(T^{-1/2})$ and $b_{ijkl} = O(T^{-1})$, so $\Delta_2 = O(T^{-1})$ and the terms left out are of order $T^{-2}$. Our main theorem in Section~\ref{s:theory} gives conditions under which $\log m_T = \log m_1 + \Delta_2 + O(T^{-2})$. This correction is well known. The three contractions in \eqref{eq:delta2} are the level-one terms of the expansion of \citet{ShunMcCullagh1995} for integrals in $d$ dimensions, enumerated by \citet[pp.~254--256]{McCullagh1987}. For $d = 1$, \eqref{eq:delta2} reduces to the correction $f^{(4)}(\hat\psi)/(8\Lambda^2) + 5f^{(3)}(\hat\psi)^2/(24\Lambda^3)$ of \citet{TierneyKadane1986}, with $\Lambda = -f''(\hat\psi)$.

The proposed estimator replaces $Z_T(\hat{\bm{\psi}})$ in \eqref{eq:m0} by an evaluation $\hat Z_T(\hat{\bm{\psi}})$ and adds the correction:
\begin{equation} \label{eq:estimator}
	\log \hat m_2 = \tfrac{d}{2}\log(2\pi) - \tfrac12\log|\bLam| + \log p_{\bm{\psi}}(\hat{\bm{\psi}}) + \log \hat Z_T(\hat{\bm{\psi}}) + \Delta_2 .
\end{equation}
We write $\hat m_1$ for the same expression without $\Delta_2$.
The subscripts distinguish the ordinary and corrected
approximations, while the hats indicate use of an evaluated
likelihood. Implementation combines an evaluator of the integrated likelihood with numerical optimization and derivative calculations.

\subsection{Evaluating the Integrated Likelihood at the Mode} \label{ss:evaluators}
The proposed approach requires an evaluator that is accurate at one point and smooth enough in $\bm{\psi}$ for the derivatives in \eqref{eq:delta2} to be computed by finite differences. Several established methods can compute the integrated likelihood. Numerical-integration filters propagate the filtering density on a grid \citep{Kitagawa1987}, and exact filters exist for special structures, such as the noncentral-gamma state dynamics of \citet{Creal2017}.\footnote{Numerical-integration filters are also developed or used by \citet{FridmanHarris1998}, \citet{Watanabe1999}, \citet*{LangrockMacDonaldZucchini2012}, \citet{BeginBoudreault2021} and \citet{Farmer2021}.} Importance samplers for the state path \citep{DurbinKoopman1997} and particle filters \citep[see][for a survey]{Creal2012} give simulation estimates, but a simulation estimate is noisy and its noise enters the finite differences directly.\footnote{Importance samplers for the state path include those of \citet{DanielssonRichard1993}, \citet{SP97}, \citet{RichardZhang2007}, \citet{McCausland12} and \citet{chan23JE}; particle filters include those of \citet*{ADH10} and \citet{PdGK12}.} We therefore use a deterministic evaluator, the numerical-integration filter on a grid. The models in this paper have scalar or bivariate states; a trivariate state is also feasible, and the cost rises by a factor equal to the number of grid points in the third dimension.

For ease of exposition, consider a scalar state, and let $\alpha_i = \alpha_{\min} + i\,\delta$, $i = 1, \ldots, N$, be equally spaced grid points. Let $\bK$ be the $N \times N$ matrix with entries $K_{ij} = p(\alpha_i \gvn \alpha_j, \bm{\theta})\,\delta$, the transition density between grid points times the trapezoidal weight $\delta$, and let $\bD_t = \diag\big(p(\by_t \gvn \alpha_1, \bm{\theta}), \ldots, p(\by_t \gvn \alpha_N, \bm{\theta})\big)$ be the diagonal matrix of the observation densities at the grid points. Further define
\[
	\ba_1 = \bD_1\,\big(p(\alpha_1 \gvn \bm{\theta}), \ldots, p(\alpha_N \gvn \bm{\theta})\big)',
	\qquad
	\ba_t = \bD_t \bK \ba_{t-1}, \quad t = 2, \ldots, T,
\]
so that $\ba_t$ approximates the joint density $p(\by_1, \ldots, \by_t, \alpha_t = \alpha_i)$ of the observations to date and the state at the grid points, and $Z_T(\bm{\theta}) \approx \delta\,\bone'\ba_T$. To avoid underflow, each $\ba_t$ is divided by $c_t = \delta\,\bone'\ba_t$, the one-step-ahead predictive density $p(\by_t \gvn \by_1, \ldots, \by_{t-1}, \bm{\theta})$, after which $\ba_t$ is the filtering density at the grid points and $\log Z_T(\bm{\theta}) \approx \sum_{t=1}^T \log c_t$. This is the numerical-integration filter of \citet{Kitagawa1987}, the grid-based filter of the nonlinear filtering literature; we call it the grid filter. To our knowledge, it has not been used for marginal likelihood evaluation. 

The filter is accurate because the integrand at each step is analytic and decays in both tails. For such an integrand, the error of the trapezoidal rule decays exponentially as the spacing $\delta$ shrinks \citep{TrefethenWeideman2014}, so a spacing of a fraction of the width of the transition and observation densities gives $\log Z_T$ to machine precision. We provide further details of the grid design and accuracy checks in Online Appendix~\ref{app:implementation}. Each step requires a multiplication by $\bK$, so the cost is linear in $T$ and in the number of grid points. 

\subsection{Mode Finding and the Approximation Error} \label{ss:budget}

We first describe how we compute the posterior mode $\hat{\bm{\psi}}$, the curvature $\bLam$ and, for $\hat m_2$, the third and fourth derivatives in \eqref{eq:delta2}. We then outline how the errors of these quantities and of the likelihood evaluation enter the proposed estimator. With the grid filter, $\log Z_T$ is deterministic and can be evaluated to an accuracy of $10^{-11}$. We use Newton--Raphson iterations in the standardized coordinates $\bu$, with gradients and Hessians computed by central differences. The negative Hessian at the mode is $\bLam$. The correction \eqref{eq:delta2} needs only the third derivatives $a_{ijk}$ and the fourth derivatives of the form $b_{iijj}$, which we compute by central differences on a sparse design in $\bu$. The whole computation requires a few hundred to several thousand evaluations of $\log Z_T$. The evaluations within each Newton iteration and within the derivative calculation are independent, so they can be run in parallel batches. With enough cores, the wall-clock cost is a few dozen evaluation times. Online Appendix~\ref{app:implementation} provides the implementation details and the evaluation counts.

Given $\hat{\bm{\psi}}$, $\bLam$ and $\Delta_2$, the error of the proposed estimator is exactly
\begin{equation} \label{eq:budget}
	\log \hat m_2 - \log m_T
	= \underbrace{\log \hat Z_T(\hat{\bm{\psi}}) - \log Z_T(\hat{\bm{\psi}})}_{\text{evaluation}}
	+ \underbrace{\Delta_2 - \log \Em\big[\e^{R(\bG)}\big]}_{\text{approximation}} .
\end{equation}
The evaluation term is zero to machine precision with the grid filter. The error of the estimator is therefore the approximation term, the error of the second-order Laplace expansion, which is deterministic given the data. We show in the next section that under regularity conditions on $f$ it is of order $T^{-2}$, while $\Delta_2$ is of order $T^{-1}$. The remaining error is therefore of the order of $\Delta_2^2$. The size of the second-order term thus indicates the accuracy of the estimate for the data at hand. We report it with every estimate. Because \eqref{eq:budget} holds for any $\hat{\bm{\psi}}$ and positive definite $\bLam$, an inexact mode or curvature enters through the approximation term and adds no separate source of error.\footnote{The finite-difference errors in the derivatives are of the order of the remainder and are reduced by extrapolation; see Online Appendix~\ref{app:implementation}.}

The second-order term also serves as a diagnostic for the reliability of the expansion. The remainder is of the order of $\Delta_2^2$, so when $|\Delta_2|$ exceeds one the indicated remainder is of the same order as the correction itself, and the expansion is not reliable. In such cases we keep $\hat{\bm{\psi}}$ and $\bLam$ and estimate the integral over the static parameters by importance sampling from a Student-$t$ proposal centered at the mode in the standardized coordinates \citep{Geweke1989}; the sampler requires no fitting step and no sampling over the states. To our knowledge, this importance-sampling estimator of the marginal likelihood has not been used for nonlinear state space models.

\section{Theoretical Results} \label{s:theory}

We provide the theoretical results behind the approximation term in \eqref{eq:budget}. Proposition~\ref{prop:cumulants} expresses the derivatives of the log integrated likelihood that enter the second-order term as cumulants under the smoothing distribution of the states. Theorem~\ref{thm:expansion} states the conditions under which the corrected approximation has an error of order $T^{-2}$.

Let $\mathcal{C}(\bm{\psi}, \bm{\alpha}) = \log p(\by \gvn \bm{\alpha}, \bm{\psi}) + \log p(\bm{\alpha} \gvn \bm{\psi})$ denote the complete-data log density, so that $Z_T(\bm{\psi}) = \int \e^{\mathcal{C}(\bm{\psi}, \bm{\alpha})}\,\di\bm{\alpha}$. Let $p(\bm{\alpha} \gvn \by, \bm{\psi})$ denote the smoothing distribution. For a fixed direction $\bw \in \mathbb{R}^d$, write $c_j = D_{\bw}^j\,\mathcal{C}(\bm{\psi}, \bm{\alpha})$ for the $j$th directional derivative in $\bm{\psi}$ at fixed $\bm{\alpha}$. Write $\Em_{\bm{\psi}}$, $\Var_{\bm{\psi}}$, $\Cov_{\bm{\psi}}$ and $\cum_{\bm{\psi}}$ for moments and joint cumulants under $p(\bm{\alpha} \gvn \by, \bm{\psi})$. The third-order joint cumulants are central moments, $\cum_{\bm{\psi}}(c_1, c_1, c_2) = \Em_{\bm{\psi}}[(c_1 - \Em_{\bm{\psi}}c_1)^2(c_2 - \Em_{\bm{\psi}}c_2)]$, and $\cum_{\bm{\psi}}(c_1, c_1, c_1, c_1) = \Em_{\bm{\psi}}[(c_1 - \Em_{\bm{\psi}}c_1)^4] - 3\Var_{\bm{\psi}}(c_1)^2$. Differentiating $Z_T$ under the integral sign gives the score identity $D_{\bw}\log Z_T = \Em_{\bm{\psi}}[c_1]$ and the observed-information identity of \citet{Louis1982}, $D_{\bw}^2\log Z_T = \Em_{\bm{\psi}}[c_2] + \Var_{\bm{\psi}}(c_1)$. The next result extends them to the third and fourth derivatives.

\begin{proposition}[Smoothing cumulants] \label{prop:cumulants}
For a fixed $\bm{\psi}$, suppose that $\mathcal{C}$ is four times continuously differentiable in its first argument, and that $\Em_{\bm{\psi}'}\exp(\sum_{j=1}^4 t_jc_j)$, with the $c_j$ evaluated at $\bm{\psi}'$, is finite for $(\mathbf{t}, \bm{\psi}')$ in a neighborhood of $(\mathbf{0}, \bm{\psi})$ and can be differentiated under the integral sign there. Then, at $\bm{\psi}$,
\begin{equation} \label{eq:d34}
\begin{split}
	D_{\bw}^3 \log Z_T &= \Em_{\bm{\psi}}[c_3] + 3\Cov_{\bm{\psi}}(c_1, c_2) + \cum_{\bm{\psi}}(c_1, c_1, c_1), \\
	D_{\bw}^4 \log Z_T &= \Em_{\bm{\psi}}[c_4] + 4\Cov_{\bm{\psi}}(c_1, c_3) + 3\Var_{\bm{\psi}}(c_2) + 6\cum_{\bm{\psi}}(c_1, c_1, c_2) \\
	&\quad + \cum_{\bm{\psi}}(c_1, c_1, c_1, c_1).
\end{split}
\end{equation}
\end{proposition}

The identities in \eqref{eq:d34} follow from the general rule for the derivative of a joint cumulant \citep{Skovgaard1986},
\begin{equation} \label{eq:dcum}
	D_{\bw}\cum_{\bm{\psi}}(g_1, \ldots, g_k) = \sum_{i=1}^k \cum_{\bm{\psi}}(g_1, \ldots, D_{\bw}g_i, \ldots, g_k) + \cum_{\bm{\psi}}(g_1, \ldots, g_k, c_1),
\end{equation}
which for $k = 1$ is $D_{\bw}\Em_{\bm{\psi}}[g] = \Em_{\bm{\psi}}[D_{\bw}g] + \Cov_{\bm{\psi}}(g, c_1)$. We prove the proposition in Online Appendix~\ref{app:proofs}. 

We assume a standard regularity condition of Laplace's method, the counterpart of a unique, interior maximum likelihood estimator with a nonsingular information matrix. In the unconstrained coordinates $\bm{\psi}$ a maximum is interior whenever it exists. 

\begin{assumption}[Mode] \label{as:mode}
The function $f$ is twice continuously differentiable, attains its maximum on $\mathbb{R}^d$ only at $\hat{\bm{\psi}}$, and $\bLam = -\nabla^2 f(\hat{\bm{\psi}})$ is positive definite.
\end{assumption}

The next assumption is stated on the set $\mathcal{N}_T = \{\hat{\bm{\psi}} + \bA\bu : \|\bu\| \leq \varrho\sqrt{T}\}$ for a fixed $\varrho > 0$.

\begin{assumption}[Derivatives] \label{as:derivatives}
The function $f$ is six times continuously differentiable on an open set containing $\mathcal{N}_T$, and there are constants $K_3, \ldots, K_6$ not depending on $T$, with $\varrho K_3 \leq 3/2$, such that, for $k = 3, \ldots, 6$,
\[
	\sup_{\bm{\psi} \in \mathcal{N}_T}\ \sup_{\|\bw\| = 1}\ \big| D^k f(\bm{\psi})[\bA\bw, \ldots, \bA\bw] \big| \leq K_k\,T^{1 - k/2} .
\]
\end{assumption}

Assumption~\ref{as:derivatives} fixes the rates at which the derivatives of $f$ grow with $T$, in the standardized coordinates in which the Laplace approximation is a standard Gaussian. Along any unit direction the $k$th standardized derivative is of order $T^{1-k/2}$: the third is $O(T^{-1/2})$, the fourth $O(T^{-1})$, and the fifth and sixth $O(T^{-3/2})$ and $O(T^{-2})$. These are the rates of a log-likelihood built from $T$ weakly dependent observations, and they make the expansion of $R$ in \eqref{eq:ratio} an expansion in powers of $T^{-1/2}$, with $C_k = O(T^{1-k/2})$, $\Delta_2 = O(T^{-1})$ and a remainder of order $T^{-2}$. The bounds are required uniformly on $\mathcal{N}_T$ because the proof uses them both on the ball in $\bu$ where it expands $f$ and on the shell between that ball and $\mathcal{N}_T$.

The rates hold when the information grows linearly in $T$. If the eigenvalues of $\bLam$ are of order $T$, then $\|\bA\|$ is of order $T^{-1/2}$, $\mathcal{N}_T$ has a fixed size in the original coordinates, and the assumption follows when the unstandardized derivatives of $f$ up to order six are of order $T$ there. For the likelihood part of $f$ these derivatives are the smoothing cumulants of Proposition~\ref{prop:cumulants} and their analogues up to order six, which are of order $T$ when the derivatives $c_j$ have moments bounded uniformly on $\mathcal{N}_T$ and the smoothing distribution forgets the past at a geometric rate, that is, when the dependence between the states at dates $s$ and $t$ decays like $\rho^{|t-s|}$ for some $\rho < 1$; the prior does not depend on $T$ and its log density is smooth, so it contributes derivatives that are bounded on $\mathcal{N}_T$. The condition $\varrho K_3 \leq 3/2$ is a choice of $\varrho$, since $K_3$ cannot increase when $\varrho$ decreases; the price is that the next assumption is then required on a smaller set.

\begin{assumption}[Concentration] \label{as:tail}
There is a constant $C_0$ not depending on $T$ such that $\int_{\mathcal{N}_T^c} \e^{f(\bm{\psi})}\,\di\bm{\psi} \leq C_0\,m_1\,T^{-2}$ for all large $T$.
\end{assumption}

Assumption~\ref{as:tail} is a weak form of posterior concentration. Under the Laplace regularity conditions of \citet*{KassTierneyKadane1990} it holds with an exponential rate in place of $T^{-2}$.

\begin{theorem}[Second-order expansion] \label{thm:expansion}
Under Assumptions~\ref{as:mode}--\ref{as:tail},
\[
	\log m_T = \log m_1 + \Delta_2 + O(T^{-2}),
\]
with $\Delta_2$ as in \eqref{eq:delta2}, and $\Delta_2 = O(T^{-1})$. The implied constant and the threshold on $T$ depend only on $d$, $\varrho$, $K_3, \ldots, K_6$ and $C_0$.
\end{theorem}

We prove the theorem in Online Appendix~\ref{app:proofs}. The proof expands $R$ to sixth order on a ball of radius $T^{1/8}$ in $\bu$, where the Gaussian expectation of the expansion gives $\Delta_2$ and all other terms are of order $T^{-2}$. It then bounds the contributions of the shell between that ball and $\mathcal{N}_T$, and of $\mathcal{N}_T^c$. The argument follows \citet*{KassTierneyKadane1990}, with the regularity conditions written in standardized coordinates. 

\section{Accuracy and Model Selection on Generated Data} \label{s:MC}

In this section we measure the accuracy of the proposed approach against quadrature benchmarks and compare it with three commonly used simulation-based estimators at equal computing time. We then generate data from three bivariate stochastic volatility models, which we also compare in the exchange-rate application, and record how often the true model has the largest estimated marginal likelihood.

\subsection{Design and Benchmark} \label{ss:design}

We consider four commonly used nonlinear or non-Gaussian state space models. The first is the stochastic volatility (SV) model in \eqref{eq:sv}. The other three vary only in their observation equations: the Student-$t$ (SV-$t$) model has $\epsilon_t \sim \distn{T}_\nu$ with unit scale, the stochastic volatility in mean (SV-M) model \citep{KoopmanHolUspensky2002, chan17} has $y_t = \lambda\e^{h_t} + \e^{h_t/2}\epsilon_t$, and the Poisson model has counts $y_t \sim \text{Poisson}(\e^{h_t})$.

Data are generated from the stochastic volatility model with $\mu = 0$ and $(\phi, \sigma)$ equal to $(0.90, 0.35)$, $(0.95, 0.25)$ or $(0.98, 0.15)$. The related models use $(\phi, \sigma) = (0.95, 0.25)$, with $\nu = 5$, $\lambda = 0.5$ and, for the Poisson model, $\mu = 2$ and $\sigma = 0.22$. We consider five sample sizes, $T \in \{500, 1{,}000, 2{,}000, 4{,}000, 8{,}000\}$, and take the first $T$ observations of each generated path, so the samples are nested. We generate ten replications for each of the six model and parameter combinations, and the five sample sizes then give 300 data sets. The replications show the size of the approximation error and its dependence on $T$. The ranges reported below are ranges over replications, not confidence intervals.

The benchmark value of $\log m_T$ for each data set comes from a nested quadrature over the three or four coordinates of $\bm{\psi}$, a trapezoidal rule in $\text{atanh}\,\phi$ with a Gauss--Hermite rule over the remaining coordinates at each node. It is very accurate for two reasons. The integrand at each node is the integrated likelihood from the grid filter, which is accurate to $10^{-11}$, so the only error is that of the outer quadrature. Since the outer quadrature is over a few smooth coordinates, refining its rules changes the value by less than $10^{-4}$ (see Online Appendix~\ref{app:implementation}). The proposed approach uses the same grid filter: the optimization routine gives $\hat{\bm{\psi}}$, $\bLam$ and the derivatives in \eqref{eq:delta2}. The residual $\log m_T - \log m_1 - \Delta_2$ reported below is the negative of the approximation term in \eqref{eq:budget}.

Finally, we consider the following priors: $\mu \sim \distn{N}(0, 10)$, $(1 + \phi)/2 \sim \distn{B}(20, 1.5)$ and $\sigma^2 \sim \distn{IG}(2.5, 0.1)$ in every model; $\lambda \sim \distn{N}(0, 1)$ in the stochastic volatility in mean model; and $\log(\nu - 2) \sim \distn{N}(2, 1)$ in the Student-$t$ model. The coordinates are $\bm{\psi} = (\mu, \text{atanh}\,\phi, \log\sigma^2)'$, extended by $\psi_4 = \lambda$ or $\psi_4 = \log(\nu - 2)$ in the two models with $d = 4$ parameters, and the Jacobian of the transform is in $p_{\bm{\psi}}$.

\FloatBarrier
\subsection{Accuracy of the Proposed Estimator} \label{ss:accuracy}

\begin{figure}[!htb]
\centering
\includegraphics[width=\textwidth]{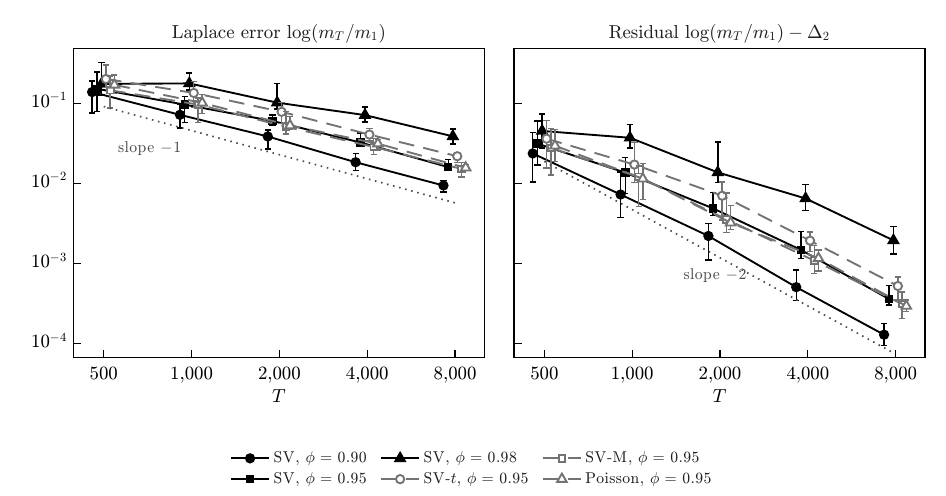}
\caption{Laplace error $\log(m_T/m_1)$ (left) and residual after the second-order term $\log(m_T/m_1) - \Delta_2$ (right) against the sample size, on logarithmic scales. Markers are medians over the replications and bars their ranges, for the SV, SV-$t$, SV-M and Poisson models. The dotted lines have slopes $-1$ and $-2$.}
\label{fig:accuracy}
\end{figure}

We first report the accuracy of the proposed estimator under the simulation design discussed above. Specifically, for each model, persistence level and sample size, Figure~\ref{fig:accuracy} plots the Laplace error $\log(m_T/m_1)$ and the residual $\log(m_T/m_1) - \Delta_2$ after the second-order term, as medians and ranges over the replications. The errors are in the logarithm, so they are approximately relative errors in $m_T$. For example, 0.01 corresponds to an error of 1\% in the marginal likelihood and in any Bayes factor computed from it. 

The left panel of Figure~\ref{fig:accuracy} shows that the Laplace approximation is reasonably accurate for every model and persistence level. Measured against the quadrature benchmark, it understates the log marginal likelihood by at most 0.33 at $T = 500$, 0.09 at $T = 4{,}000$ and 0.05 at $T = 8{,}000$. The error increases with the persistence, and the three related models behave like the stochastic volatility model at the same persistence.

The right panel shows that the second-order term removes between 64\% and 99\% of this error across the data sets. The residual falls roughly at the rate $T^{-2}$ of Theorem~\ref{thm:expansion}, at which each doubling of $T$ divides it by four. The second-order term also indicates the size of the residual: in all 300 data sets the residual lies between 0.5 and 7 times $\Delta_2^2$, with median 1.5, so we use about $2\Delta_2^2$ as the indication of the remaining error. The second-order term $\Delta_2$ itself is at most 0.25 in these data sets, so the indicated error is at most about 0.12, and no data set comes near the threshold of one at which the expansion stops being reliable.

\subsection{Comparison with Simulation-Based Estimators} \label{ss:comparison}

We now compare the proposed estimator with three commonly used marginal likelihood estimators on data sets simulated from the stochastic volatility model in Section~\ref{ss:design}: importance sampling with the cross-entropy method of \citet{CE15}, Chib's method \citep{chib95}, and the modified harmonic mean of \citet{GelfandDey1994} with the weighting density of \citet{Geweke1999}. All three require posterior draws from MCMC.

The three estimators share one posterior sampler, the auxiliary mixture sampler of \citet*{KSC98}. We obtain 20,000 posterior draws after a burn-in period of 5,000. The cross-entropy method fits a Gaussian proposal in $\bm{\psi}$ to these draws. It then evaluates the integrated likelihood at 10,000 draws from this proposal. Chib's method computes the posterior ordinate at the posterior mean of the draws, from the full run and two reduced runs. The Gelfand--Dey estimator evaluates the integrated likelihood at each of the 20,000 posterior draws, as \citet{CG15} recommend for latent variable models. Because the auxiliary mixture sampler targets the mixture model, these two estimators reweight the draws to the exact model as in \citet*{KSC98}; the importance weights of the cross-entropy method correct for the mixture approximation directly. All four estimators use the grid filter for the integrated likelihood, so the comparison isolates the integration over the parameters.

\begin{figure}[!htb]
\centering
\includegraphics[width=\textwidth]{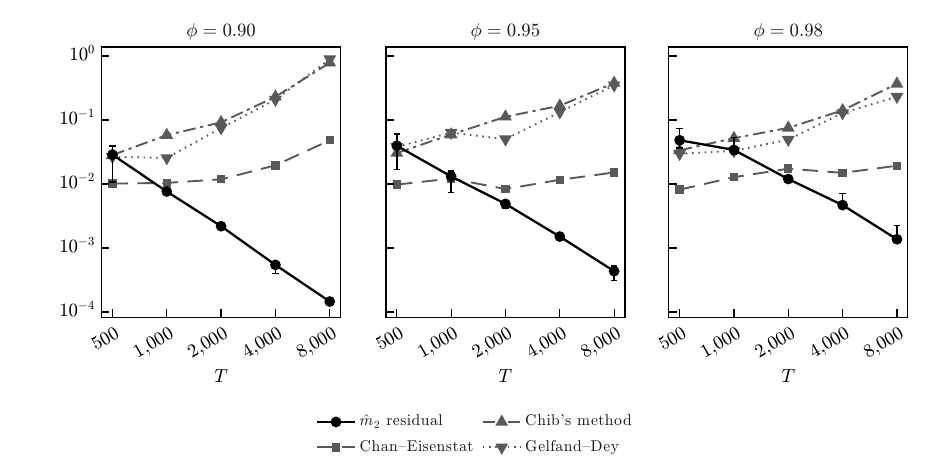}
\caption{Error of the four marginal likelihood estimators (in log scale) at a computing budget of one minute on one core, against the sample size, for the stochastic volatility model at three persistence levels. For $\hat m_2$ the error is the residual, as median and range over the replications; for the other estimators it is the numerical standard error scaled to the budget.}
\label{fig:compare}
\end{figure}

To compare the estimators at equal cost, we rescale the numerical standard errors of the three simulation-based estimators to a computing budget of one minute on one core. We measure each standard error with the fixed numbers of draws given above, together with the computing time of each estimator on one core.\footnote{The computations of this section are done in $\matlab$. The timings are single-thread times on a laptop computer with an Intel Core Ultra 7 255U processor.} Since the number of draws is proportional to the computing time, a numerical standard error is inversely proportional to the square root of the computing time. Figure~\ref{fig:compare} plots these rescaled standard errors next to the error of the proposed estimator, by sample size and persistence level. 

At this budget the error of the proposed estimator falls as $T$ grows. The errors of Chib's method and the Gelfand--Dey estimator rise steeply with $T$, and that of the cross-entropy method rises by a factor of about two to five. The proposed estimator is therefore the most accurate of the four from $T = 1{,}000$ at $\phi = 0.90$ and from $T = 2{,}000$ at $\phi \geq 0.95$. At $T = 4{,}000$ its residual is 2 to 47 times smaller than the standard error of the cross-entropy method, the most accurate of the three simulation-based estimators, and at $T = 8{,}000$ it is 9 to 400 times smaller. 

The simulation-based estimators lose accuracy as $T$ grows for two reasons. First, each MCMC sweep and each likelihood evaluation takes longer at larger $T$. A fixed budget therefore allows fewer draws as $T$ grows, and from $T = 4{,}000$ the three runs of Chib's method alone exceed it. The proposed estimator requires a number of likelihood evaluations that does not grow with $T$, and it takes 2.5 to 63 seconds, within the budget at $T \leq 4{,}000$ and about the budget at $T = 8{,}000$. Second, the reweighting of Chib's method and the Gelfand--Dey estimator to the exact model keeps a smaller share of the draws as $T$ grows: at $T = 4{,}000$ the effective sample size is 0.1 to 3 percent of the draws, and at $T = 8{,}000$ it is below 0.5 percent. Their standard errors therefore grow with $T$ even at a fixed number of draws, and at $T \geq 4{,}000$ their realized errors exceed these standard errors by up to a factor of six (see Online Appendix~\ref{app:implementation}). The cross-entropy method requires no reweighting. Its standard error at a fixed number of draws falls with $T$ as the posterior becomes closer to Gaussian, which offsets much of the rising cost of each draw.

Overall, these results show that the proposed estimator is accurate for long time series at a small computing cost. Its error is roughly proportional to $T^{-2}$, and it takes about a minute or less on one core at every sample size. The simulation-based estimators require longer runs at larger $T$ to reach a given accuracy. The advantage of the proposed estimator therefore grows with $T$.

\subsection{Model Selection for Bivariate Stochastic Volatility} \label{ss:mcfx}

The Monte Carlo experiments above use scalar states, for which a quadrature benchmark exists. The models of the exchange-rate application have bivariate states and seven or eight static parameters, for which no benchmark is available. We assess how often the proposed estimator selects the true model by marginal likelihood. We compare three of the bivariate stochastic volatility models of \citet{YuMeyer2006}, members of the multivariate stochastic volatility family reviewed by \citet*{AsaiMcAleerYu2006}. These models are costly to estimate by MCMC. No auxiliary mixture samplers in the tradition of \citet*{KSC98} have been developed for them. \citet{YuMeyer2006} estimate these models using single-move samplers in WinBUGS---which \citet*{AsaiMcAleerYu2006} note can be numerically inefficient---and compare them by the deviance information criterion.

We write a pair of returns as $\by_t = (y_{1t}, y_{2t})'$ with $y_{it} = \e^{h_{it}/2}\epsilon_{it}$, $i = 1, 2$. In the constant-correlation model (CC), the two log-volatilities follow independent AR(1) processes as in \eqref{eq:sv}, and $\bm{\epsilon}_t = (\epsilon_{1t}, \epsilon_{2t})' \sim \distn{N}(\bm{0}, \bm{P})$, where $\bm{P}$ is a correlation matrix with off-diagonal element $\rho$. CC has a two-dimensional state and seven static parameters. Each of the other two models changes one feature of CC. The Granger-causality model (GC) adds the lagged first log-volatility to the second transition, $h_{2,t+1} = \mu_2 + \phi_2(h_{2t} - \mu_2) + \phi_{21}(h_{1t} - \mu_1) + \sigma_2\eta_{2t}$. The heavy-tailed model (CC-$t$) replaces the Gaussian $\bm{\epsilon}_t$ by a bivariate Student-$t$ with scale matrix $\bm{P}$ and $\nu > 2$ degrees of freedom, so that $\Var(\bm{\epsilon}_t) = \nu\bm{P}/(\nu - 2)$. Each has eight static parameters.

We use the same priors as in Section~\ref{ss:design} for $(\mu_i, \phi_i, \sigma_i^2)$. The other priors are $\rho \sim \distn{U}(-1, 1)$, $\phi_{21} \sim \distn{N}(0, 1)$ and $\log(\nu - 2) \sim \distn{N}(2, 1)$. The coordinates of CC are $\bm{\psi} = (\mu_1, \text{atanh}\,\phi_1, \log\sigma_1^2, \mu_2, \text{atanh}\,\phi_2, \log\sigma_2^2, \text{atanh}\,\rho)'$. GC adds $\psi_8 = \phi_{21}$ and CC-$t$ adds $\psi_8 = \log(\nu - 2)$. We generate data from each of the three models, with the parameters at their posterior modes for the exchange-rate data of Section~\ref{s:app}. For each model we generate 50 series of $T = 1{,}000$ observations, a quarter of the exchange-rate sample, which makes the models harder to distinguish. We fit all three models to every series, with $\log\hat m_2$ from the grid filter as the log marginal likelihood. 

\begin{figure}[!htb]
\centering
\includegraphics[width= 0.9\textwidth]{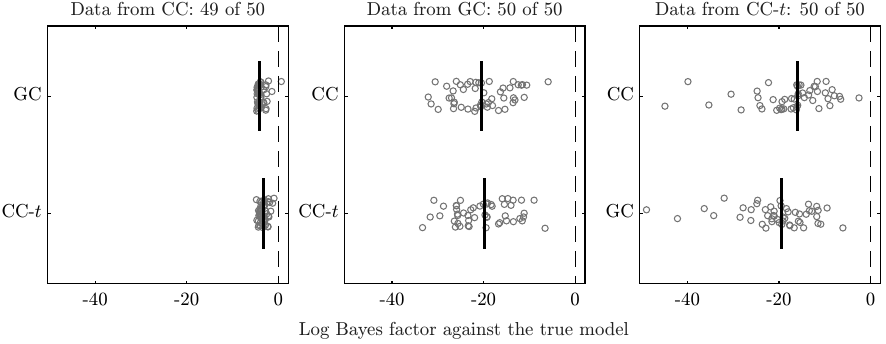}
\caption{Log Bayes factors of the other two models against the true model on 50 data sets of $T = 1{,}000$ observations generated from CC (left), GC (middle) and CC-$t$ (right). Each dot is one series, jittered vertically, and the bar is the median; values below zero favor the true model.}
\label{fig:mcfx}
\end{figure}

For each series, Figure~\ref{fig:mcfx} plots the log Bayes factors of the other two models against the true model. The true model is selected in 149 of the 150 series. The exception is a series generated from CC, on which GC is selected by 0.66---posterior odds of about two to one. Both other models include CC as a special or limiting case: GC when $\phi_{21} = 0$ and CC-$t$ as $\nu \to \infty$. On data from CC, the more flexible models fit as well as CC. They have only one additional static parameter, so the penalty for their complexity is modest. The median log Bayes factors against CC are therefore only $-4.1$ for GC and $-3.3$ for CC-$t$. In contrast, when the true model has a feature that the other two lack, it is selected in every series. For data from GC and CC-$t$, the median log Bayes factors against the true model range from $-16.0$ to $-20.5$, and the 95th percentiles are $-6.8$ or below.

These results show that the proposed estimator reliably identifies the true model among the three bivariate specifications. The one exception reflects a property of the Bayes factor: when the true model is nested in a more flexible one, the evidence for the restricted model accumulates only slowly with the sample size.

\FloatBarrier
\section{Application: Bivariate SV for Exchange Rates} \label{s:app}

This section applies the proposed approach to three bivariate stochastic volatility models of \citet{YuMeyer2006} for two exchange rates. The models have bivariate states and seven or eight static parameters. To our knowledge, the marginal likelihood has never been reported for these models. Posterior sampling for them is itself difficult---no auxiliary mixture sampler exists, and the existing samplers are single-move and numerically inefficient \citep*{AsaiMcAleerYu2006}. We describe the data and the models, then present the Bayes factors and the computing cost.

\subsection{Data and Models} \label{ss:appmodels}

We use daily log returns on the Australian and New Zealand dollars against the US dollar, constructed from the European Central Bank's daily euro reference rates. The sample spans 4,000 trading days from February 16, 2011, to October 2, 2026. Returns are expressed in percent and demeaned over the full ECB series, from January 1999 to October 2026. The two series have standard deviations of approximately 0.65 and 0.68 percent and a sample correlation of 0.83.

We compare the three bivariate stochastic volatility models introduced in Section~\ref{ss:mcfx}. They share the observation equation $y_{it}=\e^{h_{it}/2}\epsilon_{it}$ and first-order autoregressive log-volatility dynamics. The baseline constant-correlation model (CC) assumes Gaussian return shocks with correlation $\rho$. Each alternative extends the baseline in one direction. The Granger-causality model (GC) allows lagged AUD log-volatility to enter the NZD log-volatility equation. The heavy-tailed model (CC-$t$) replaces the Gaussian return shocks with a bivariate Student-$t$ distribution with $\nu$ degrees of freedom.

The integrated likelihood of each model is evaluated on a product grid in $(h_{1t}, h_{2t})$, and the derivatives in \eqref{eq:delta2} come from the sparse design outlined in Section~\ref{ss:budget}.\footnote{The transition kernel of the grid filter is the product of two one-dimensional banded kernels, or a block-banded kernel for GC. One evaluation takes between 4 and 14 seconds on one core at this $T$, and the evaluations of each optimization step run in parallel on 16 workers, in $\matlab$ on a desktop computer with an Intel Core i7-14700 processor. The sparse design has 770 points in seven dimensions and 1,168 in eight.} Online Appendix~\ref{app:implementation} provides the details of the optimization settings.

\subsection{Bayes Factors and Computing Cost} \label{ss:appresults}
% Merged the two paragraph into one so that we explain our results and also compare to their results, which reads more smoothly.
Both extensions receive strong support relative to the constant-correlation baseline, and heavy-tailed observations provide the larger improvement. Table~\ref{tab:app} shows log Bayes factors against CC of 110.3 for CC-$t$ and 97.7 for GC. The heavy-tailed model is the best supported model of the three, whereas \citet{YuMeyer2006}, who use weekly returns, rank the heavy-tailed model below CC by the deviance information criterion (DIC). This is not surprising as daily returns typically have heavier tails than weekly returns. The parameter estimates help explain these gains. CC-$t$ accommodates heavy tails with a degrees-of-freedom parameter of $\nu = 7.9$, while leaving the volatility parameters close to their baseline estimates. In addition, the data show a strong volatility spillover from the Australian dollar to the New Zealand dollar ($\phi_{21} = 0.83$), in line with their results. Once this spillover is taken into account, the coefficient $\phi_2$ on the own lag of NZD log-volatility falls to 0.003.

 %GC changes the dynamics of NZD volatility more substantially: the coefficient $\phi_2$ on its own lag falls to 0.003, and the coefficient $\phi_{21}$ on lagged AUD log-volatility is 0.83.

\begin{table}[!htb]
\caption{Bivariate stochastic volatility models for the AUD/USD and NZD/USD returns, $T = 4{,}000$. The parameter values are those at the posterior mode. The column $\Delta_2$ is the second-order term, and the column $\log\tbf$ gives the log Bayes factor against CC, whose $\log\hat m_2$ is $-5{,}474.86$. Evaluations are the number of likelihood evaluations, and minutes are the wall-clock time on 16 workers.}
\label{tab:app}
\centering
\footnotesize\setlength{\tabcolsep}{3pt}
\begin{tabular}{lcccccccrr}
\toprule
Model & $\phi_1$ & $\phi_2$ & $\rho$ & $\phi_{21}$ & $\nu$ & $\Delta_2$ & $\log\tbf$ & Evaluations & Minutes \\
\midrule
CC & 0.982 & 0.984 & 0.836 & -- & -- & 0.15 & 0 & 1,976 & 12 \\
GC & 0.977 & 0.003 & 0.845 & 0.829 & -- & 0.08 & 97.7 & 4,330 & 129 \\
CC-$t$ & 0.985 & 0.987 & 0.837 & -- & 7.9 & 0.18 & 110.3 & 3,266 & 32 \\
\bottomrule
\end{tabular}

\end{table}

%\citet{YuMeyer2006} compare these specifications on 519 weekly returns from 1994 to 2003 by the deviance information criterion (DIC). Their results agree with ours on volatility spillovers: the coefficient on lagged AUD log-volatility is 0.49, and the DIC favors GC over CC. They differ on heavy tails: their posterior mean of $\nu$ is 23, and the DIC ranks the heavy-tailed model below CC. Daily returns typically have heavier tails than weekly returns, and in our sample heavy tails are the best supported extension.

The second-order term $\Delta_2$ is between 0.08 and 0.18 for the three models, well below the threshold of one; the remaining approximation error indicated by $2\Delta_2^2$ is below 0.07 for each log marginal likelihood. The log Bayes factors in Table~\ref{tab:app} are therefore accurate to about 0.1, and the ranking is unaffected by the approximation error.

The computational cost is modest even for a large sample of $T = 4{,}000$. The three models require between 2,000 and 4,300 likelihood evaluations each, taking from 12 minutes to about two hours on 16 workers. MCMC-based alternatives would require posterior simulation for each specification, with additional reduced runs for Chib's method. For a Gelfand--Dey estimator based on the integrated likelihood, evaluating the likelihood at tens of thousands of posterior draws would itself be costly. At 4 to 14 seconds per evaluation, 20,000 evaluations would take about one to three days on a single core, compared with 2 to 17 hours of single-core time for the optimization underlying our approximation.

\section{Application: SV for Intraday Returns} \label{s:app2}

Our second application examines whether extreme intraday returns are better explained by heavy-tailed observations, jumps, or short-lived movements in volatility. We compare seven stochastic volatility models, four with a scalar state and three with two volatility factors, on $T = 19{,}266$ five-minute returns on the SPDR S\&P 500 ETF (SPY) over 2025. The models have seven to eleven static parameters. We first describe the data and the models, report the Bayes factors and the computing cost, and then examine how the ranking of the models develops through the year.

\subsection{Data and Models} \label{ss:app2models}
The data are five-minute log returns in percent on the SPY in the regular session, 9:30 to 16:00, on the 247 full trading days of 2025. The price of each interval is its last trade and the price at 9:30 is the first trade of the day, so there are 78 returns per day and no overnight return. The returns are demeaned over the sample.\footnote{The trades are from the TAQ millisecond consolidated trades file on Wharton Research Data Services (WRDS), 201.6 million in the year. We drop trades whose condition codes mark them as out of sequence, canceled, corrected or outside regular hours, and we exclude the three early-close days (July 3, November 28 and December 24).}

The kurtosis of the returns is 153. Without the four most volatile days, April 7 to 10, the autocorrelation of the squared returns is 0.23 at lag 1, 0.12 at lag 20 and still 0.11 at lag 78, one trading day later. It halves within 20 lags and then barely declines, which suggests a fast-moving volatility component in addition to a persistent one. Volatility also varies with the time of day: the root mean square return is 0.13 percent in the first five minutes of the session, 0.08 percent between 12:00 and 13:00 and 0.10 percent in the last five minutes, which suggests a recurring time-of-day component could be beneficial. The extreme returns also motivate specifications with heavy tails and with jumps.

We compare seven specifications that accommodate these features in different ways. The baseline model is the stochastic volatility model in \eqref{eq:sv}, denoted SV. The Student-$t$ specification, SV-$t$, replaces its Gaussian return shock with $\epsilon_t \sim \distn{T}_\nu$ with unit scale. The remaining models introduce return jumps, volatility jumps, or a second volatility factor. These extensions build on specifications considered by \citet{ChernovGallantGhyselsTauchen2003} for daily stock index returns.

The return-jump model, SVJ, adds an occasional shock to the observation equation, as in \citet*{ChibNardariShephard2002}:
\[
y_t = \e^{h_t/2}\epsilon_t + b_t\zeta_t,
\]
where the jump indicator $b_t$ equals one with probability $\omega$ and zero otherwise, the jump size is $\zeta_t \sim \distn{N}(0, \sigma_\zeta^2)$, and both are independent over time and of the other shocks. Given $h_t$, the return is a mixture of two normal distributions with mean zero, variances $\e^{h_t}$ and $\e^{h_t} + \sigma_\zeta^2$, and weights $1 - \omega$ and $\omega$. The state remains the scalar $h_t$.

In the fourth model (SVJV) the same jump also enters the log-volatility, as in the model of \citet*{ErakerJohannesPolson2003} with jumps in returns and volatility:
\[
y_t = \e^{h_t/2}\epsilon_t + b_t\zeta_t, \qquad h_t = \mu + \phi(h_{t-1} - \mu) + \sigma\eta_t + b_t\xi_t, \qquad \xi_t \sim \distn{N}(\mu_\xi, \sigma_\xi^2),
\]
so that a jump in the return arrives together with a step in the log-volatility, which then decays at the rate $\phi$. Given $h_{t-1}$, the pair $(y_t, h_t)$ is a mixture of two components with weights $1 - \omega$ and $\omega$: the no-jump transition with the observation density $\distn{N}(0, \e^{h_t})$, and the transition shifted by $\xi_t$ with the observation density $\distn{N}(0, \e^{h_t} + \sigma_\zeta^2)$. The state is again the scalar $h_t$, with $h_1$ from the stationary distribution of the recursion without jumps. Through the shared jump the observation density depends on $(h_{t-1}, h_t)$, and the grid filter sums over the two components at each date.

The last three models split the log-volatility into a persistent and a transitory factor:
\[
y_t = \e^{(h_{1t} + h_{2t})/2}\epsilon_t, \qquad h_{1t} = \mu + \phi_1(h_{1,t-1} - \mu) + \sigma_1\eta_{1t}, \qquad h_{2t} = \phi_2 h_{2,t-1} + \sigma_2\eta_{2t},
\]
where $\eta_{1t}$ and $\eta_{2t}$ are independent standard normal and both factors start from their stationary distributions. The restriction $\phi_2 < \phi_1$ makes $h_{1t}$ the persistent factor. The two-factor stochastic volatility model (TF) has a Gaussian $\epsilon_t$, the two-factor Student-$t$ model (TF-$t$) has $\epsilon_t \sim \distn{T}_\nu$ with unit scale, and the two-factor jump model (TFJ) adds the jump $b_t\zeta_t$ of the third model to the return. Their state is two-dimensional, and the grid filter evaluates the integrated likelihood on a product grid.

All seven models account for the pronounced time-of-day pattern in intraday volatility \citep{AndersenBollerslev1997}. We add to the log-volatility in the observation equation a deterministic component $\kappa(\tau_t)$ with coefficients $\kappa_1, \ldots, \kappa_4$. The component accounts for the recurring daily pattern, so that the latent factors describe only the fluctuations around it. The coefficients are estimated jointly with the dynamics, as in \citet{StroudJohannes2014}, so the Bayes factors do not depend on a pattern estimated beforehand.\footnote{The component has the first two harmonics of the trading day, $\kappa(\tau_t) = \kappa_1\cos(2\pi\tau_t) + \kappa_2\sin(2\pi\tau_t) + \kappa_3\cos(4\pi\tau_t) + \kappa_4\sin(4\pi\tau_t)$, where $\tau_t = j_t/78$ and $j_t \in \{1, \ldots, 78\}$ indexes the five-minute intervals of the trading day. It multiplies the variance of the non-jump return component by $\e^{\kappa(\tau_t)}$ and enters the observation density only, so the filters are unchanged.}

We use the same priors on $\mu$, $\phi$, $\sigma^2$ and $\nu$ as in Section~\ref{ss:design}, with $(\phi_1, \sigma_1^2)$ in the role of $(\phi, \sigma^2)$ in the two-factor models. The jump probability has the prior $\omega \sim \distn{B}(2, 100)$, with mean 0.02, and the jump variance has the prior $\sigma_\zeta^2 \sim \distn{IG}(2.5, 0.1)$. The volatility jump has the priors $\mu_\xi \sim \distn{N}(0, 4)$ and $\sigma_\xi^2 \sim \distn{IG}(2.5, 0.1)$. In the two-factor models, $\phi_2 \gvn \phi_1 \sim \distn{U}(-1, \phi_1)$ and $\sigma_2^2 \sim \distn{IG}(2.5, 0.1)$. The coefficients of the time-of-day term have the priors $\kappa_1, \ldots, \kappa_4 \sim \distn{N}(0, 1)$. We use the same coordinates as in Section~\ref{ss:design}, extended by $(\text{logit}\,\omega, \log\sigma_\zeta^2)$ in the jump models, by $(\mu_\xi, \log\sigma_\xi^2)$ in the volatility-jump model and by $(w, \log\sigma_2^2)$ in the two-factor models, where $w = \log(\text{atanh}\,\phi_1 - \text{atanh}\,\phi_2)$ with $\phi_2 < \phi_1$. The coefficients $\kappa_1, \ldots, \kappa_4$ are coordinates in every model, without transformation.

\subsection{Bayes Factors and Computing Cost} \label{ss:app2results}

For all seven models, the marginal likelihood estimator is $\log\hat m_2$ from the grid filter, with the derivatives in \eqref{eq:delta2} from the sparse design of Section~\ref{ss:budget}. Table~\ref{tab:app2} reports, for each model, the volatility and jump parameters at the posterior mode, the second-order term, the log Bayes factor against SV and the computing cost.\footnote{The coefficients of the time-of-day term differ by less than 0.02 across the seven models and are omitted from the table; in SV they are $(\kappa_1, \kappa_2, \kappa_3, \kappa_4) = (0.30, 0.36, 0.16, 0.23)$. The fitted pattern rises from 2.9 times its minimum at the open to 3.5 times at about 10:00, falls to the minimum at 14:30 and rises again to 2.7 times at the close. The computations of this section are done in $\matlab$ on a desktop computer with an Intel Core i9-12900KS processor. The evaluations of each optimization step run on eight workers.} Heavy tails and return jumps improve the log marginal likelihood over SV by 25.1 and 12.3, respectively. Larger gains come from more flexible volatility dynamics: SVJV and TF improve on SV by 68.9 and 73.2. The parameter estimates help explain why these extensions are useful. SV has a single persistence parameter of 0.987 (an implied half-life of 4.4 hours). TF separates a persistent component with $\phi_1=0.996$ (a half-life of about two trading days) from a transitory component with $\phi_2=0.834$ (a half-life of nineteen minutes). The innovation variance of the transitory component is ten times that of the persistent component, so large volatility movements can occur and dissipate quickly.

\begin{table}[!htb]
\caption{Stochastic volatility models for the five-minute SPY returns, $T = 19{,}266$. The parameter values are those at the posterior mode; in the two-factor models, $\phi$ and $\sigma^2$ belong to the persistent factor and $\phi_2$ and $\sigma_2^2$ to the transitory factor. The column $\Delta_2$ is the second-order term, and the column $\log\tbf$ gives the log Bayes factor against SV, whose $\log\hat m_2$ is $25{,}352.52$. Evaluations are the number of likelihood evaluations, and minutes are the wall-clock time on eight workers.}
\label{tab:app2}
\centering
\footnotesize\setlength{\tabcolsep}{3pt}
% generated by make_table.m on 2026-10-19 from spy5taq_raw, T = 19266, K = 2
\begin{tabular}{lcccccccccccrr}
\toprule
Model & $\phi$ & $\sigma^2$ & $\phi_2$ & $\sigma_2^2$ & $\nu$ & $\omega$ & $\sigma_\zeta^2$ & $\mu_\xi$ & $\sigma_\xi^2$ & $\Delta_2$ & $\log\tbf$ & Evaluations & Minutes \\
\midrule
SV & 0.987 & 0.040 & -- & -- & -- & -- & -- & -- & -- & 0.02 & 0 & 2,794 & 3 \\
SV-$t$ & 0.990 & 0.028 & -- & -- & 15.9 & -- & -- & -- & -- & 0.04 & 25.1 & 3,538 & 4 \\
SVJ & 0.988 & 0.034 & -- & -- & -- & 0.0035 & 0.047 & -- & -- & 0.07 & 12.3 & 4,668 & 8 \\
SVJV & 0.989 & 0.019 & -- & -- & -- & 0.0068 & 0.032 & 1.33 & 0.054 & 0.31 & 68.9 & 9,044 & 19 \\
TF & 0.996 & 0.010 & 0.834 & 0.098 & -- & -- & -- & -- & -- & 0.07 & 73.2 & 5,342 & 57 \\
TF-$t$ & 0.996 & 0.009 & 0.854 & 0.083 & 55.3 & -- & -- & -- & -- & 0.09 & 69.4 & 7,230 & 138 \\
TFJ & 0.996 & 0.010 & 0.847 & 0.087 & -- & 0.0010 & 0.077 & -- & -- & 0.15 & 71.8 & 8,550 & 208 \\
\bottomrule
\end{tabular}

\end{table}

SVJV also accommodates abrupt volatility increases, with an estimated jump probability implying approximately 131 expected jump events over the year and a mean log-volatility jump of 1.33. Their effects, however, decay with a half-life of about five hours. The log Bayes factor of 4.3 in favor of TF over SVJV suggests that the shorter-lived transitory component provides a better account of these movements.

Once the transitory volatility factor is included, adding heavy tails or return jumps reduces the log marginal likelihood by about 3.7 and 1.4, respectively. The parameter estimates also suggest less need for either feature: the degrees-of-freedom parameter increases from 15.9 in SV-$t$ to 55.3 in TF-$t$, while the jump probability falls from 0.0035 in SVJ to 0.0010 in TFJ, corresponding to approximately 67 and 20 expected jumps over the year. The second-order term $\Delta_2$ ranges from 0.02 to 0.31 across the seven models, well below the threshold of one. The remaining approximation error indicated by $2\Delta_2^2$ is therefore below 0.2 for every log marginal likelihood, and it does not affect any of the comparisons above.

\subsection{Model Comparison Through the Year}\label{ss:app2recursive}

The four models TF, TFJ, TF-$t$ and SVJV have log Bayes factors against SV within 4.3 of one another. Over 19{,}266 returns, such differences may reflect small gains accumulated steadily through the sample or large gains concentrated in a few episodes such as the tariff shock of April, when jumps are known to matter \citep{ErakerJohannesPolson2003, MaheuMcCurdy2004}. We therefore examine whether TF is favored consistently through the year or whether the best model changes over time. For $s < t$, $\log m_t - \log m_s$ is the predictive log likelihood of the observations between $s$ and $t$, and its difference between two models is their predictive log Bayes factor over that period \citep{GewekeAmisano2010}. We fit each model on the sample up to the end of each month from February through December. The sample to the end of February serves as the training sample, and the differences between consecutive fits are the predictive log Bayes factors of the ten months from March to December.\footnote{The returns are demeaned over the full sample, so the exercise decomposes the full-sample Bayes factors retrospectively and is not a real-time comparison.}

The proposed approach makes this exercise computationally feasible. It requires 77 marginal likelihoods, seven models at eleven month ends. Each is a mode search with a few thousand likelihood evaluations, started from the mode of the previous month end, and no posterior sampler is needed. A simulation-based estimator would require a posterior simulation for every model at every month end.

\begin{figure}[!htb]
	\centering
	\includegraphics[width=0.88\textwidth]{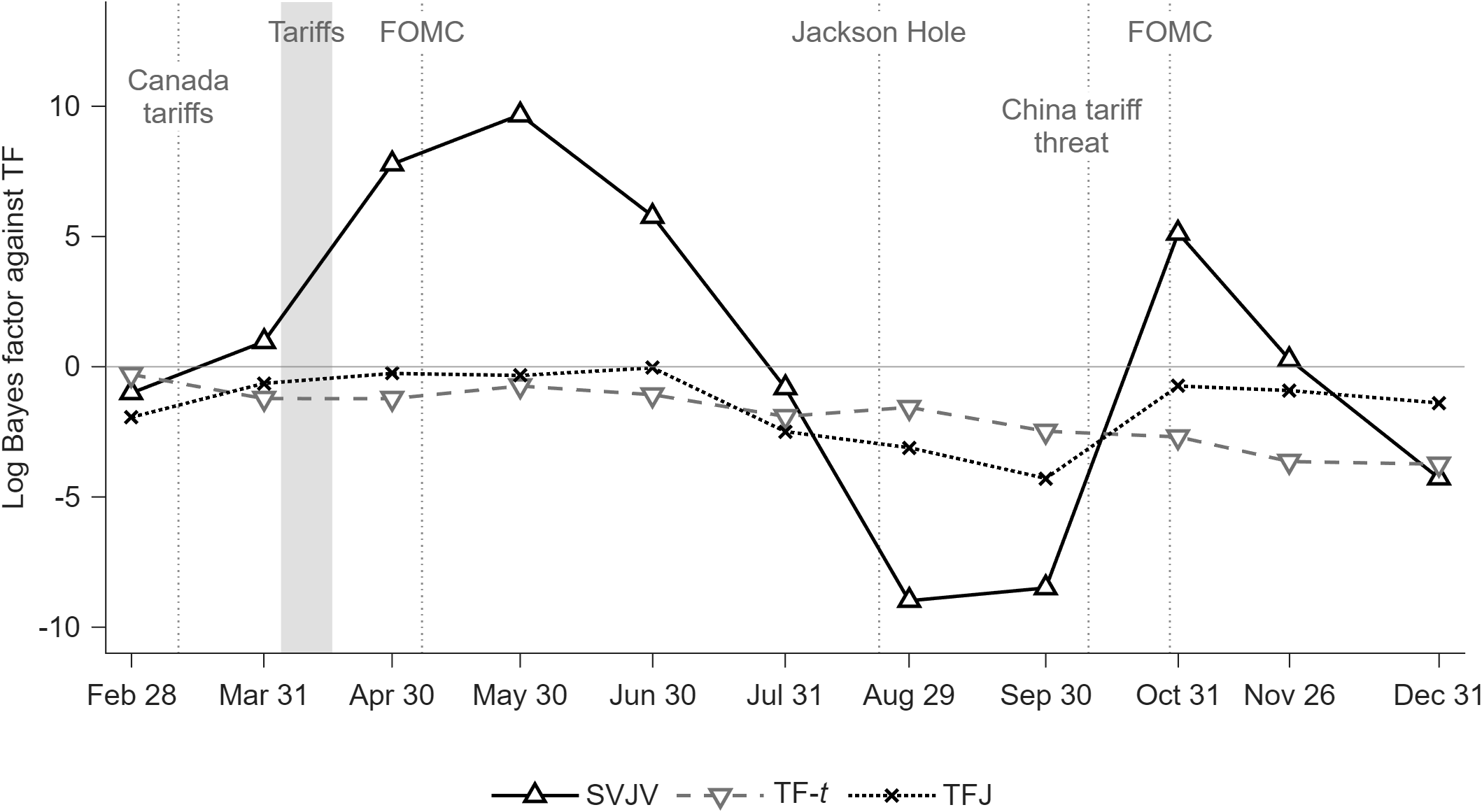}
	\caption{Log Bayes factors of SVJV, TF-$t$ and TFJ against TF on the sample up to each month end of 2025, from the end of the training sample on February 28; above zero the model is favored over TF on the sample so far. The marked events are the days with a five-minute return on the SPY above 0.5 percent in absolute value during the 9:30 -- 16:00 trading window: the tariffs on Canadian steel and aluminum (March 11), the tariff episode of April 4 to 16, the FOMC announcements of May 7 and October 29, the Jackson Hole speech (August 22) and the China tariff threat (October 10).}
	\label{fig:recursive}
\end{figure}

Figure~\ref{fig:recursive} shows the log Bayes factors of SVJV, TF-$t$ and TFJ against TF, the best model in the full sample, on the sample up to each month end; a value above zero means that the model is favored over TF on the sample available at that date. The paths start at the log Bayes factors on the training sample, $-1.0$, $-0.3$ and $-1.9$, and end at the full-sample values of Table~\ref{tab:app2}, about $-4.3$, $-3.7$ and $-1.4$. We also mark the events on the days when a five-minute return on the SPY exceeds 0.5 percent in absolute value: the tariffs on Canadian steel and aluminum (March 11), the tariff episode of April 4 to 16, the FOMC announcements of May 7 and October 29, the Jackson Hole speech (August 22) and the China tariff threat (October 10).

The gains of SVJV over TF mostly coincide with the marked events, except the Jackson Hole speech. The strongest evidence in its favor comes from two months: its log Bayes factor against TF rises by 6.8 in April, the month of the tariff episode, and by 13.6 in October, which contains the China tariff threat and an FOMC announcement. It also rises by 2.0 in March and by 1.9 in May, the months of the tariffs on Canadian steel and aluminum and of another FOMC announcement.  For all three models the net change over the ten months is small relative to these monthly movements, and TF remains favored in the full sample, as it was at the end of the training sample.

The 77 fits also illustrate how the second-order term serves as a diagnostic. It is at most 0.45 at 74 of the fits and exceeds one at three, each time because the variance of the jump sizes is informed by only 25 to 110 jumps. At those three fits we use the mode and curvature already computed to construct the importance-sampling proposal, as described in Section~\ref{ss:budget}, and the resulting estimate has a numerical standard error of 0.04. Online Appendix~\ref{app:diagnostic} reports the monthly log Bayes factors and the details of the three fits.

\FloatBarrier
\section{Concluding Remarks} \label{s:conclusion}

We propose a deterministic estimator of the marginal likelihood of nonlinear state space models. A grid filter integrates out the states, and a Laplace approximation with its second-order term integrates over the static parameters. No posterior sampler is required, and the second-order term indicates the remaining error. For stochastic volatility models the estimator is accurate to within 0.01 in the log marginal likelihood at $T = 4{,}000$, and at equal computing time it is more accurate than three commonly used simulation-based estimators from $T = 2{,}000$. In the applications, heavy tails are the best supported extension for the exchange rates, and a transitory volatility factor is decisively supported for the intraday returns.

There are two interesting directions for future research. The first concerns models with states of four or more dimensions. The grid filter is then too costly, and a particle filter would have to evaluate the integrated likelihood, so the mode and the curvature would be found from noisy values. The second is real-time use. Updating the mode and curvature at every observation would take only a few likelihood evaluations, so the predictive likelihood $m_{T+1}/m_T$ and the posterior model weights could be updated each period. This would give real-time Bayesian model averaging and density forecast comparison for nonlinear state space models without a posterior sampler for any model.

\subsubsection*{Declaration of Generative AI}

We acknowledge the use of Claude Code and Codex for editing, brainstorming and checking the code. A previous version of the paper was reviewed by Refine.ink, and its suggestions are incorporated in the current draft. All AI-assisted suggestions were carefully reviewed, and we take full responsibility for the final content and accuracy of this paper.

\subsubsection*{Data Availability Statement}

The exchange rate data are the European Central Bank's daily euro reference rates, which are publicly available from the ECB Data Portal at \url{https://data.ecb.europa.eu/data/datasets/EXR}. The intraday data are from the NYSE Trade and Quote (TAQ) database on Wharton Research Data Services. They are proprietary and cannot be redistributed. The code and the exchange rate data will be provided as online supplementary material.

\begin{singlespace}
\footnotesize
\setlength{\bibsep}{0pt}
\bibliographystyle{econometrica}
\bibliography{ml-Laplace}

@article{RichardZhang2007,
  author  = {Richard, Jean-Fran\c{c}ois and Zhang, Wei},
  title   = {Efficient High-Dimensional Importance Sampling},
  journal = {Journal of Econometrics},
  volume  = {141},
  number  = {2},
  pages   = {1385--1411},
  year    = {2007}
}

@article{DanielssonRichard1993,
  author  = {Dan\'{i}elsson, J\'{o}n and Richard, Jean-Fran\c{c}ois},
  title   = {Accelerated {G}aussian Importance Sampler with Application to Dynamic Latent Variable Models},
  journal = {Journal of Applied Econometrics},
  volume  = {8}, number = {S1}, pages = {S153--S173}, year = {1993}
}

@article{DurbinKoopman1997,
  author  = {Durbin, James and Koopman, Siem Jan},
  title   = {Monte {C}arlo Maximum Likelihood Estimation for Non-{G}aussian State Space Models},
  journal = {Biometrika}, volume = {84}, number = {3}, pages = {669--684}, year = {1997}
}

@article{Farmer2021,
  author  = {Farmer, Leland E.},
  title   = {The Discretization Filter: A Simple Way to Estimate Nonlinear State Space Models},
  journal = {Quantitative Economics},
  volume  = {12},
  number  = {1},
  pages   = {41--76},
  year    = {2021}
}

@article{chan23JE,
  author  = {J. C. C. Chan},
  title   = {Comparing Stochastic Volatility Specifications for Large {B}ayesian {VAR}s},
  journal = {Journal of Econometrics},
  volume  = {235},
  number  = {2},
  pages   = {1419--1446},
  year    = {2023}
}

@article{CYZ26,
  author  = {J. C. C. Chan and X. Yu and W. Zhang},
  title   = {{B}ayesian Model Comparison for Large {B}ayesian {VAR}s after the {COVID}-19 Pandemic},
  journal = {Journal of Econometrics},
  volume  = {256}, pages = {106072}, year = {2026}
}

@article{ADH10,
  title={Particle {M}arkov Chain {M}onte {C}arlo Methods},
  author={Andrieu, C. and Doucet, A. and Holenstein, R.},
  journal={Journal of the Royal Statistical Society Series B},
  volume={72},
  number={3},
  pages={269--342},
  year={2010},
}

@article{PdGK12,
  title={On Some Properties of {M}arkov Chain {M}onte {C}arlo Simulation Methods Based on the Particle Filter},
  author={Pitt, M. K. and dos Santos Silva, R. and Giordani, P. and Kohn, R.},
  journal={Journal of Econometrics},
  volume={171},
  number={2},
  pages={134--151},
  year={2012},
  publisher={Elsevier}
}

@ARTICLE{chib95,
author={S. Chib},
title={Marginal Likelihood from the \uppercase{G}ibbs Output},
journal={Journal of the American Statistical Association},
volume={90},
number={432},
pages={1313--1321},
year={1995}
}

@ARTICLE{CJ01,
author={S. Chib and I. Jeliazkov},
title={Marginal Likelihood from the \uppercase{M}etropolis-\uppercase{H}astings Output},
journal={Journal of the American Statistical Association},
volume={96},
number={453},
pages={270--281},
year={2001}
}

@article {CE15,
author = {Chan, J. C. C. and Eisenstat, E. },
title = {Marginal Likelihood Estimation with the Cross-Entropy Method},
journal = {Econometric Reviews},
year = {2015},
volume = {34},
number = {3},
pages = {256-285},
}

@article{KSC98,
	title = "Stochastic Volatility: Likelihood Inference and Comparison with \textsc{ARCH} Models",
	author = "S. Kim and N. Shephard and S. Chib",
	journal = "Review of Economic Studies",
	pages = "361--393",
	volume = "65",
	number = "3",
	year = "1998"
}

@article{SP97,
  author = {Shephard, N. and M. K. Pitt},
  title = {Likelihood Analysis of Non-{G}aussian Measurement Time Series},
  journal = {Biometrika},
  volume = {84},
  number = {3},
  pages = {653--667},
  year = {1997}
}

@article{McCausland12,
title = {The {HESSIAN} Method: Highly Efficient Simulation Smoothing, in a Nutshell},
journal = {Journal of Econometrics},
volume = {168},
number = {2},
year = {2012},
author = {W. J. McCausland},
pages = {189--206},
}

@unpublished{Chan2026,
  author = {Chan, J. C. C.},
  title = {Exact Rejection Sampling for Non-{G}aussian State Space Models},
  note = {arXiv:2608.21619},
  year = {2026},
}

@article{TrefethenWeideman2014,
  author = {Trefethen, L. N. and Weideman, J. A. C.},
  title = {The Exponentially Convergent Trapezoidal Rule},
  journal = {{SIAM} Review},
  year = {2014},
  volume = {56},
  pages = {385--458},
}

@article{AndersenBollerslev1997,
  author = {Andersen, T. G. and Bollerslev, T.},
  title = {Intraday Periodicity and Volatility Persistence in Financial Markets},
  journal = {Journal of Empirical Finance},
  year = {1997},
  volume = {4},
  pages = {115--158},
}

@article{ChernovGallantGhyselsTauchen2003,
  author = {Chernov, M. and Gallant, A. R. and Ghysels, E. and Tauchen, G.},
  title = {Alternative Models for Stock Price Dynamics},
  journal = {Journal of Econometrics},
  year = {2003},
  volume = {116},
  pages = {225--257},
}

@article{TierneyKadane1986,
  author = {Tierney, L. and Kadane, J. B.},
  title = {Accurate Approximations for Posterior Moments and Marginal Densities},
  journal = {Journal of the American Statistical Association},
  volume = {81},
  number = {393},
  pages = {82--86},
  year = {1986},
}

@incollection{KassTierneyKadane1990,
  author = {Kass, R. E. and Tierney, L. and Kadane, J. B.},
  title = {The Validity of Posterior Expansions Based on {L}aplace's Method},
  booktitle = {Bayesian and Likelihood Methods in Statistics and Econometrics},
  editor = {Geisser, S. and Hodges, J. S. and Press, S. J. and Zellner, A.},
  publisher = {North-Holland},
  address = {Amsterdam},
  pages = {473--488},
  year = {1990},
}

@article{ShunMcCullagh1995,
  author = {Shun, Z. and McCullagh, P.},
  title = {Laplace Approximation of High Dimensional Integrals},
  journal = {Journal of the Royal Statistical Society Series B},
  volume = {57},
  number = {4},
  pages = {749--760},
  year = {1995},
}

@article{Skovgaard1986,
  author = {Skovgaard, I. M.},
  title = {A Note on the Differentiation of Cumulants of Log Likelihood Derivatives},
  journal = {International Statistical Review},
  volume = {54},
  number = {1},
  pages = {29--32},
  year = {1986}
}

@article{Louis1982,
  author = {Louis, T. A.},
  title = {Finding the Observed Information Matrix When Using the {EM} Algorithm},
  journal = {Journal of the Royal Statistical Society Series B},
  volume = {44},
  number = {2},
  pages = {226--233},
  year = {1982},
}

@article{Kitagawa1987,
  author = {Kitagawa, G.},
  title = {Non-{G}aussian State-Space Modeling of Nonstationary Time Series},
  journal = {Journal of the American Statistical Association},
  volume = {82},
  number = {400},
  pages = {1032--1041},
  year = {1987},
}

@article{FridmanHarris1998,
  author = {Fridman, M. and Harris, L.},
  title = {A Maximum Likelihood Approach for Non-{G}aussian Stochastic Volatility Models},
  journal = {Journal of Business and Economic Statistics},
  volume = {16},
  number = {3},
  pages = {284--291},
  year = {1998},
}

@article{LangrockMacDonaldZucchini2012,
  author = {Langrock, R. and MacDonald, I. L. and Zucchini, W.},
  title = {Some Nonstandard Stochastic Volatility Models and Their Estimation Using Structured Hidden {M}arkov Models},
  journal = {Journal of Empirical Finance},
  volume = {19},
  number = {1},
  pages = {147--161},
  year = {2012},
}

@article{BeginBoudreault2021,
  author = {B\'egin, J.-F. and Boudreault, M.},
  title = {Likelihood Evaluation of Jump-Diffusion Models Using Deterministic Nonlinear Filters},
  journal = {Journal of Computational and Graphical Statistics},
  volume = {30},
  number = {2},
  pages = {452--466},
  year = {2021},
}

@article{ChopinJacobPapaspiliopoulos2013,
  author = {Chopin, N. and Jacob, P. E. and Papaspiliopoulos, O.},
  title = {{SMC}$^2$: An Efficient Algorithm for Sequential Analysis of State Space Models},
  journal = {Journal of the Royal Statistical Society Series B},
  volume = {75},
  number = {3},
  pages = {397--426},
  year = {2013},
}

@article{MengWong1996,
  author = {Meng, X.-L. and Wong, W. H.},
  title = {Simulating Ratios of Normalizing Constants via a Simple Identity: A Theoretical Exploration},
  journal = {Statistica Sinica},
  volume = {6},
  number = {4},
  pages = {831--860},
  year = {1996},
}

@article{Watanabe1999,
  author = {Watanabe, T.},
  title = {A Non-Linear Filtering Approach to Stochastic Volatility Models with an Application to Daily Stock Returns},
  journal = {Journal of Applied Econometrics},
  volume = {14},
  number = {2},
  pages = {101--121},
  year = {1999},
}

@article{Geweke1989,
  author = {Geweke, J.},
  title = {Bayesian Inference in Econometric Models Using {M}onte {C}arlo Integration},
  journal = {Econometrica},
  volume = {57},
  number = {6},
  pages = {1317--1339},
  year = {1989},
}

@article{YuMeyer2006,
  author = {Yu, J. and Meyer, R.},
  title = {Multivariate Stochastic Volatility Models: {B}ayesian Estimation and Model Comparison},
  journal = {Econometric Reviews},
  volume = {25},
  number = {2--3},
  pages = {361--384},
  year = {2006},
}

@book{BurdenFaires2011,
  author    = {Burden, R. L. and Faires, J. D.},
  title     = {Numerical Analysis},
  edition   = {9th},
  publisher = {Brooks/Cole},
  address   = {Boston},
  year      = {2011}
}

@book{McCullagh1987,
  author    = {McCullagh, P.},
  title     = {Tensor Methods in Statistics},
  publisher = {Chapman and Hall},
  address   = {London},
  year      = {1987}
}

@article{RueMartinoChopin2009,
  author  = {Rue, H. and Martino, S. and Chopin, N.},
  title   = {Approximate {B}ayesian Inference for Latent {G}aussian Models by Using Integrated Nested {L}aplace Approximations},
  journal = {Journal of the Royal Statistical Society Series B},
  volume  = {71},
  number  = {2},
  pages   = {319--392},
  year    = {2009}
}

@article{RuliSartoriVentura2016,
  author  = {Ruli, E. and Sartori, N. and Ventura, L.},
  title   = {Improved {L}aplace Approximation for Marginal Likelihoods},
  journal = {Electronic Journal of Statistics},
  volume  = {10},
  number  = {2},
  pages   = {3986--4009},
  year    = {2016}
}

@article{Creal2012,
  author  = {Creal, D.},
  title   = {A Survey of Sequential {M}onte {C}arlo Methods for Economics and Finance},
  journal = {Econometric Reviews},
  volume  = {31},
  number  = {3},
  pages   = {245--296},
  year    = {2012}
}

@article{Creal2017,
  author  = {Creal, D.},
  title   = {A Class of Non-{G}aussian State Space Models with Exact Likelihood Inference},
  journal = {Journal of Business and Economic Statistics},
  volume  = {35},
  number  = {4},
  pages   = {585--597},
  year    = {2017}
}

@article{CJ09,
  author  = {Chan, J. C. C. and Jeliazkov, I.},
  title   = {Efficient Simulation and Integrated Likelihood Estimation in State Space Models},
  journal = {International Journal of Mathematical Modelling and Numerical Optimisation},
  volume  = {1},
  number  = {1--2},
  pages   = {101--120},
  year    = {2009}
}

@article{chan17,
  author  = {Chan, J. C. C.},
  title   = {The Stochastic Volatility in Mean Model with Time-Varying Parameters: An Application to Inflation Modeling},
  journal = {Journal of Business and Economic Statistics},
  volume  = {35},
  number  = {1},
  pages   = {17--28},
  year    = {2017}
}

@article{CG15,
  author  = {Chan, J. C. C. and Grant, A. L.},
  title   = {Pitfalls of Estimating the Marginal Likelihood Using the Modified Harmonic Mean},
  journal = {Economics Letters},
  volume  = {131},
  pages   = {29--33},
  year    = {2015}
}

@article{KoopmanHolUspensky2002,
  author  = {Koopman, S. J. and Hol Uspensky, E.},
  title   = {The Stochastic Volatility in Mean Model: Empirical Evidence from International Stock Markets},
  journal = {Journal of Applied Econometrics},
  volume  = {17},
  number  = {6},
  pages   = {667--689},
  year    = {2002}
}

@article{AsaiMcAleerYu2006,
  author  = {Asai, M. and McAleer, M. and Yu, J.},
  title   = {Multivariate Stochastic Volatility: A Review},
  journal = {Econometric Reviews},
  volume  = {25},
  number  = {2--3},
  pages   = {145--175},
  year    = {2006}
}

@article{LiYuZeng2020,
  author  = {Li, Y. and Yu, J. and Zeng, T.},
  title   = {Deviance Information Criterion for Latent Variable Models and Misspecified Models},
  journal = {Journal of Econometrics},
  volume  = {216},
  number  = {2},
  pages   = {450--493},
  year    = {2020}
}

@article{GelfandDey1994,
  author  = {Gelfand, A. E. and Dey, D. K.},
  title   = {{B}ayesian Model Choice: Asymptotics and Exact Calculations},
  journal = {Journal of the Royal Statistical Society Series B},
  volume  = {56},
  number  = {3},
  pages   = {501--514},
  year    = {1994}
}

@article{Geweke1999,
  author  = {Geweke, J.},
  title   = {Using Simulation Methods for {B}ayesian Econometric Models: Inference, Development, and Communication},
  journal = {Econometric Reviews},
  volume  = {18},
  number  = {1},
  pages   = {1--73},
  year    = {1999}
}

@book{BarndorffNielsenCox1989,
  author    = {Barndorff-Nielsen, O. E. and Cox, D. R.},
  title     = {Asymptotic Techniques for Use in Statistics},
  publisher = {Chapman and Hall},
  address   = {London},
  year      = {1989}
}

@article{ChibNardariShephard2002,
  author = {Chib, S. and Nardari, F. and Shephard, N.},
  title = {{M}arkov Chain {M}onte {C}arlo Methods for Stochastic Volatility Models},
  journal = {Journal of Econometrics},
  volume = {108},
  number = {2},
  pages = {281--316},
  year = {2002},
}

@article{ErakerJohannesPolson2003,
  author = {Eraker, B. and Johannes, M. and Polson, N.},
  title = {The Impact of Jumps in Volatility and Returns},
  journal = {Journal of Finance},
  volume = {58},
  number = {3},
  pages = {1269--1300},
  year = {2003},
}

@article{StroudJohannes2014,
  author = {Stroud, J. R. and Johannes, M. S.},
  title = {{B}ayesian Modeling and Forecasting of 24-Hour High-Frequency Volatility},
  journal = {Journal of the American Statistical Association},
  volume = {109},
  number = {508},
  pages = {1368--1384},
  year = {2014},
}

@article{MaheuMcCurdy2004,
	author  = {Maheu, John M. and McCurdy, Thomas H.},
	title   = {News Arrival, Jump Dynamics, and Volatility Components
	for Individual Stock Returns},
	journal = {The Journal of Finance},
	year    = {2004},
	volume  = {59},
	number  = {2},
	pages   = {755--793},
	doi     = {10.1111/j.1540-6261.2004.00648.x}
}

@article{GewekeAmisano2010,
	author  = {Geweke, J. and Amisano, G.},
	title   = {Comparing and evaluating {B}ayesian predictive distributions of asset returns},
	journal = {International Journal of Forecasting},
	year    = {2010},
	volume  = {26},
	number  = {2},
	pages   = {216--230}
}

@article{LangeVanOsVanDijk2026,
  author  = {Lange, R.-J. and van Os, B. and van Dijk, D.},
  title   = {Implicit Score-Driven Filters for Time-Varying Parameter Models},
  journal = {Journal of Econometrics},
  volume  = {255}, pages = {106251}, year = {2026}
}

@unpublished{HansenTong2026,
  author = {Hansen, P. R. and Tong, C.},
  title  = {{T}weedie's Formula and Score-Driven Updating},
  note   = {arXiv:2605.15902},
  year   = {2026},
}
\end{singlespace}
\clearpage
\appendix
\makeatletter
\renewcommand{\@seccntformat}[1]{\csname #1prefix\endcsname\csname the#1\endcsname\quad}
\renewcommand\section{\@startsection{section}{1}{\z@}{-3.5ex \@plus -1ex \@minus -.2ex}{2.3ex \@plus .2ex}{\large\bfseries}} % appendix headings one size smaller, so that the longest fits on one line
\newcommand{\sectionprefix}{Online Appendix~}
\newcommand{\subsectionprefix}{}
\makeatother

\section{Proofs} \label{app:proofs}

\textbf{Proof of Proposition~\ref{prop:cumulants}.} Write $Z = Z_T(\bm{\psi})$ and $\pi(\bm{\alpha}) = \e^{\mathcal{C}(\bm{\psi}, \bm{\alpha})}/Z$. For smooth $g_1, \ldots, g_k$ let $\mathcal{K}(\mathbf{t}) = \log\Em_{\bm{\psi}}\exp(\sum_i t_ig_i)$ be their joint cumulant generating function, so that $\cum_{\bm{\psi}}(g_1, \ldots, g_k) = \partial_{t_1}\cdots\partial_{t_k}\mathcal{K}(\bm{0})$. Differentiating under the integral sign,
\[
	D_{\bw}\mathcal{K}(\mathbf{t}) = \tilde\Em_{\mathbf{t}}\Big[\sum_i t_i D_{\bw}g_i + c_1\Big] - \Em_{\bm{\psi}}[c_1],
\]
where $\tilde\Em_{\mathbf{t}}$ is the expectation under the density proportional to $\exp(\sum_it_ig_i)\pi(\bm{\alpha})$, and $\tilde\Em_{\mathbf{t}}[q] = \partial_s\log\Em_{\bm{\psi}}\exp(\sum_it_ig_i + sq)|_{s=0}$ for any $q$. Applying $\partial_{t_1}\cdots\partial_{t_k}$ at $\mathbf{t} = \bm{0}$ gives \eqref{eq:dcum}. The term $t_iD_{\bw}g_i$ contributes $\cum_{\bm{\psi}}(g_1, \ldots, D_{\bw}g_i, \ldots, g_k)$, and the term $c_1$ contributes $\cum_{\bm{\psi}}(g_1, \ldots, g_k, c_1)$. Differentiating $Z$ under the integral sign gives $D_{\bw}\log Z = \Em_{\bm{\psi}}[c_1]$ directly. Since $D_{\bw}c_j = c_{j+1}$, repeated application of \eqref{eq:dcum} gives
\begin{align*}
	D_{\bw}\Em_{\bm{\psi}}[c_1] &= \Em_{\bm{\psi}}[c_2] + \Var_{\bm{\psi}}(c_1), \\
	D_{\bw}\Var_{\bm{\psi}}(c_1) &= 2\Cov_{\bm{\psi}}(c_1, c_2) + \cum_{\bm{\psi}}(c_1, c_1, c_1), \\
	D_{\bw}\Cov_{\bm{\psi}}(c_1, c_2) &= \Var_{\bm{\psi}}(c_2) + \Cov_{\bm{\psi}}(c_1, c_3) + \cum_{\bm{\psi}}(c_1, c_1, c_2), \\
	D_{\bw}\cum_{\bm{\psi}}(c_1, c_1, c_1) &= 3\cum_{\bm{\psi}}(c_1, c_1, c_2) + \cum_{\bm{\psi}}(c_1, c_1, c_1, c_1),
\end{align*}
and \eqref{eq:d34} follows by summing. $\square$

\textbf{Proof of Theorem~\ref{thm:expansion}.} Let $\varphi$ denote the $\distn{N}(\bm{0}, \bI_d)$ density, let $\rho_T = T^{1/8}$, and write $\beta_k = K_k T^{1-k/2}$. Take $T$ large enough that $\rho_T \leq \varrho\sqrt{T}$ and $\sum_{k=3}^6\beta_k\rho_T^k/k! \leq \log 2$. By~\eqref{eq:ratio}, $m_T/m_1 = I_1 + I_2 + I_3$, where $I_1$ is the integral of $\e^{R}\varphi$ over the ball $\|\bu\| \leq \rho_T$, $I_2$ is the integral over the shell $\rho_T < \|\bu\| \leq \varrho\sqrt{T}$, and $I_3 = m_1^{-1}\int_{\mathcal{N}_T^c}\e^{f}$. Assumption~\ref{as:tail} gives $I_3 \leq C_0T^{-2}$.

For $\|\bu\| \leq \varrho\sqrt{T}$, the segment from $\bm{0}$ to $\bu$ lies in $\mathcal{N}_T$ because $\mathcal{N}_T$ is convex. Taylor's theorem along this segment gives $R(\bu) = D^3 f(\bm{\xi})[\bA\bu, \bA\bu, \bA\bu]/6$ for some $\bm{\xi} \in \mathcal{N}_T$, so $|R(\bu)| \leq \beta_3\|\bu\|^3/6 \leq (\varrho K_3/6)\|\bu\|^2 \leq \tfrac14\|\bu\|^2$ by Assumption~\ref{as:derivatives}. Hence $\e^{R(\bu)}\varphi(\bu) \leq (2\pi)^{-d/2}\e^{-\|\bu\|^2/4}$ on $\mathcal{N}_T$, and
\[
	I_2 \leq (2\pi)^{-d/2}\int_{\|\bu\| > \rho_T}\e^{-\|\bu\|^2/4}\,\di\bu = \frac{2^{d/2}\,\Gamma(d/2, \rho_T^2/4)}{\Gamma(d/2)} \leq M_d\,\rho_T^{d-2}\,\e^{-\rho_T^2/4}
\]
for a constant $M_d$, where $\Gamma(\cdot, \cdot)$ is the upper incomplete gamma function. This bound is $o(T^{-2})$.

On the ball, Taylor's theorem to sixth order gives $R = C_3 + C_4 + C_5 + R_6$. Here $|C_k(\bu)| \leq \beta_k\|\bu\|^k/k!$ for every $\bu$ by Assumption~\ref{as:derivatives} at $\bm{\psi} = \hat{\bm{\psi}}$, and $|R_6(\bu)| \leq \beta_6\|\bu\|^6/720$ on the ball by the same assumption on $\mathcal{N}_T$. On the ball $|R(\bu)| \leq \sum_{k=3}^6\beta_k\rho_T^k/k! = O(T^{-1/8})$, and $\e^{R} = \sum_{j=0}^{4}R^j/j! + E_5$ with $|E_5| \leq |R|^5\e^{|R|}/120 \leq |R|^5/60$. Each monomial in the expansion of $\sum_{j \leq 4}R^j/j!$, and the bound on $E_5$, is dominated by a constant times a product of factors $\beta_{k_i}\|\bu\|^{k_i}$. The integral of such a product against $\varphi$ over the ball is at most the corresponding Gaussian moment over $\mathbb{R}^d$, which is finite, times $\prod_i\beta_{k_i} = O(T^{\sum_i(1 - k_i/2)})$. The terms of order $T^{-1/2}$ and $T^{-3/2}$ are $C_3$, $C_5$, $C_3C_4$ and $C_3^3/6$. They are odd functions of $\bu$ and integrate to zero over the ball. The terms of order $T^{-1}$ are $C_4$ and $\tfrac12 C_3^2$. Every other term, including $E_5$, is of order $T^{-2}$ or smaller. Therefore
\[
	I_1 = \int_{\|\bu\| \leq \rho_T}\big\{1 + C_4(\bu) + \tfrac12 C_3(\bu)^2\big\}\varphi(\bu)\,\di\bu + O(T^{-2}),
\]
and extending the integral to $\mathbb{R}^d$ changes it by at most a constant times $(1 + \beta_4\rho_T^4 + \beta_3^2\rho_T^6)\rho_T^{d-2}\e^{-\rho_T^2/2}$, which is $o(T^{-2})$. The Gaussian moments give $\Em[C_4(\bG)] = \tfrac18\sum_{i,j}b_{iijj}$, because $\Em[G_iG_jG_kG_l]$ is the sum over the three pairings of the indices. They also give $\tfrac12\Em[C_3(\bG)^2] = \tfrac1{12}\sum_{i,j,k}a_{ijk}^2 + \tfrac18\sum_i(\sum_j a_{ijj})^2$, because $\Em[G_iG_jG_kG_lG_mG_n]$ is the sum over the fifteen pairings, of which six pair each index of one cubic factor with an index of the other and nine pair one index within each factor. Hence $m_T/m_1 = 1 + \Delta_2 + O(T^{-2})$. By polarization, i.e., the identity that expresses a symmetric multilinear form through its diagonal values, Assumption~\ref{as:derivatives} at $\bm{\psi} = \hat{\bm{\psi}}$ bounds every entry of the standardized tensors, so $\Delta_2 = O(T^{-1})$, and $\log(1 + \Delta_2 + r) = \Delta_2 + O(T^{-2})$ for $r = O(T^{-2})$. All constants depend only on $d$, $\varrho$, $K_3, \ldots, K_6$ and $C_0$. $\square$

\section{Implementation Details} \label{app:implementation}

This appendix gives the implementation details behind the results in Sections~\ref{s:MC} to \ref{s:app2}. We first describe the grid filter for the scalar-state models and the quadrature benchmark, then the mode search and the sparse design for the derivatives in $\Delta_2$. The last two subsections cover the MCMC-based marginal likelihood estimators of Section~\ref{ss:comparison} and the parameter transformations with their Jacobians.

\subsection{Grid Filter and Benchmark} \label{app:evaluators}

We first describe the grid filter for the models in Section~\ref{s:MC}, whose state is scalar with transition $h_t = \mu + \phi(h_{t-1} - \mu) + \sigma\eta_t$. At a given $\bm{\theta}$, Newton's method with the banded precision matrix gives the smoothing mode $\hat\bh$ and the marginal standard deviations $\hat s_t$ of the Gaussian approximation to $p(\bh \gvn \by, \bm{\theta})$. The grid has spacing $\delta = \min(\sigma/2, 0.05)$ and runs from $\min_t\hat h_t - 12\max_t\hat s_t - 6\sigma$ to $\max_t\hat h_t + 12\max_t\hat s_t + 6\sigma$. With a spacing of at most half the transition standard deviation, the aliasing error of the trapezoidal rule for the Gaussian kernel is at most of order $\e^{-8\pi^2}$. The product grids of the bivariate models in Section~\ref{s:app} have spacing $\min(\sigma_i/2, 0.1)$ in each state, and those of the two-factor models in Section~\ref{s:app2} have spacing $0.6\sigma_i$.

One evaluation of $\log Z_T$ takes between 0.01 seconds at $T = 500$ and 0.19 seconds at $T = 8{,}000$ on one core. For a linear Gaussian model, whose integrated likelihood is available in closed form, the filter agrees with the exact value to $3 \times 10^{-12}$. On the 300 data sets in Section~\ref{s:MC}, it also agrees with an independent simulation estimate of $\log Z_T(\hat{\bm{\psi}})$, from an importance sampler for the state path with the bounded-weight proposal of \citet{Chan2026}; the largest difference is 0.001.

The benchmark value of $\log m_T$ in Section~\ref{s:MC} is a nested quadrature in $\bm{\psi}$. The outer rule is a trapezoidal rule in $z = \text{atanh}\,\phi$ with step $0.5$ posterior standard deviations, extended until the integrand falls below $10^{-9}$ of its maximum. The integrand is analytic in $z$ with Gaussian-like tails, so the trapezoidal rule converges exponentially and the error of the outer rule is far smaller than the truncation at $10^{-9}$. The inner rule is a Gauss--Hermite rule over the remaining coordinates with 8 points per dimension (6 when the inner integral is three-dimensional), centered at the conditional mode and scaled by the conditional curvature. Refining both rules changes $\log m_T$ by at most $4.2 \times 10^{-5}$ on the 240 data sets with $T \leq 4{,}000$, and by at most $3 \times 10^{-9}$ on the nine stochastic volatility data sets of the comparison at $T = 8{,}000$. For the Student-$t$ model at $T \leq 2{,}000$, where the conditional posterior of $\psi_4$ has a long right tail, the inner rule in $\psi_4$ is a trapezoidal rule.

\subsection{Mode Search and Sparse Design} \label{app:optimization}

The mode search uses Newton steps in the standardized coordinates $\bu = \bA^{-1}(\bm{\psi} - \hat{\bm{\psi}})$, with the gradient and Hessian from central differences.\footnote{Fourth-order central differences of step $0.1$, with Richardson extrapolation for the mixed second derivatives. The nine scales are $0.25, 0.5, 1, 1.5, 2, 3, 4, 6, 8$.} Each step is evaluated at nine scales in one parallel batch and the best is taken, because for some models the full Newton step is too short along a curved ridge of the posterior, as for GC. Where the Hessian is not negative definite, the step follows the gradient. In Section~\ref{s:app2} it takes the modified Newton direction described below instead. The iteration stops when the standardized step is below $10^{-7}$ in the experiments and $10^{-6}$ in the applications. For the bivariate models in Section~\ref{s:app}, the search starts from a point constructed from univariate fits. Each volatility process starts at the posterior mode of a univariate stochastic volatility model fitted to its own series, and the correlation at the sample correlation.

The derivatives in \eqref{eq:delta2} come from central differences of $R(\bu)$ at $\bu = \mathbf{0}$ on a sparse design with step $\ell = 0.25$: the points $\pm \ell\be_i$ and $\pm 2\ell\be_i$ on each axis give $a_{iii}$ and $b_{iiii}$, the four points $\pm \ell\be_i \pm \ell\be_j$ in each coordinate plane give $a_{iij}$, $a_{ijj}$ and $b_{iijj}$, and the eight points $\pm \ell\be_i \pm \ell\be_j \pm \ell\be_k$ in each coordinate cube give $a_{ijk}$. The design is sparse because \eqref{eq:delta2} needs only the third derivatives and the fourth derivatives with paired indices, so its points lie on the coordinate axes, planes and cubes, $6d + 8\binom{d}{2} + 16\binom{d}{3}$ in all, against $5^d$ for a full central-difference grid. These differences have an error of order $\ell^2$, which under Assumption~\ref{as:derivatives} changes $\Delta_2$ by $O(T^{-2})$, the order of the remainder in Theorem~\ref{thm:expansion}; repeating them at step $2\ell$ and taking $(4D_\ell - D_{2\ell})/3$ for each derivative $D$ removes this leading term \citep[Section~4.2]{BurdenFaires2011}. In the experiments the optimization requires between 258 and 760 evaluations per data set, of which the design takes 58 for $d = 3$ and 136 for $d = 4$.

In the two-factor stochastic volatility models in Section~\ref{s:app2} the Hessian is indefinite over most of the path from the start, along a ridge in $(\text{atanh}\,\phi_1, w)$, where gradient steps advance slowly. The search therefore takes a modified Newton step there: the eigenvalues of the standardized Hessian are replaced by the negatives of their absolute values, clipped to $[-4, -0.25]$, and the step is capped at two standardized units. The three two-factor searches converge in 10, 11 and 10 iterations.

\subsection{MCMC-Based Marginal Likelihood Estimators} \label{app:mcmc}

Chib's method, the Gelfand--Dey estimator and the cross-entropy method require posterior draws. For the stochastic volatility model in Section~\ref{ss:comparison}, we obtain them with the auxiliary mixture sampler of \citet*{KSC98} in conjunction with the precision-based draw of the volatility path \citep{CJ09}. The sampler uses the seven-component mixture, Gibbs steps for $\mu$ and $\sigma^2$, and the independence Metropolis--Hastings step of \citet*{KSC98} for $\phi$. The full run starts from $\phi = 0.95$, $\sigma^2 = 0.1$ and $\mu$ equal to the logarithm of the sample variance.

Chib's method evaluates the posterior ordinate at the posterior mean of the full run: the ordinate of $\mu$ comes from the full run, that of $\sigma^2$ from a reduced run with $\mu$ fixed, and that of $\phi$ from a second reduced run, normalized by quadrature. The Gelfand--Dey estimator uses the truncated normal of \citet{Geweke1999} at the 95\% ellipsoid in $\bm{\psi}$, and the cross-entropy method a Gaussian in $\bm{\psi}$ with the mean and covariance of the draws of the full run. Because the auxiliary mixture sampler targets the mixture approximation, Chib's method and the Gelfand--Dey estimator reweight each draw to the exact model. The weight is $\prod_{t=1}^T p(\varepsilon_t^*)/q(\varepsilon_t^*)$, where $\varepsilon_t^* = \log y_t^2 - h_t$, $p$ is its exact $\log\chi^2_1$ density and $q$ the seven-component mixture. The cross-entropy method uses the integrated likelihood of the exact model and needs no reweighting.

Numerical standard errors are computed by batch means with 20 batches. The computing times in Section~\ref{ss:comparison} are single-thread times on an idle core, extrapolated from the time of one sweep of each run and of one likelihood evaluation at each $T$ and $\phi$. At $T = 4{,}000$, where the reweighting keeps 0.1 to 3 percent of the draws, the largest errors are 0.76 for Chib's method and 0.13 for the Gelfand--Dey estimator, 6.2 and 4.5 times their standard errors. At $T = 8{,}000$ the reweighting keeps at most 0.5 percent of the draws, and the largest errors are 1.45 and 0.37, with standard errors of up to 1.7 and 0.11. The errors of the cross-entropy method are at most 0.020 at every $T$.

\subsection{Parameter Transformations and Jacobians} \label{app:jacobians}

This subsection records the transformations that map the parameters of each model to the unconstrained coordinates $\bm{\psi}$, and the terms that they add to $\log p_{\bm{\psi}}$. For a scalar parameter $\theta$ with prior density $p(\theta)$ and coordinate $\psi = g(\theta)$, the prior density of $\psi$ is $p(g^{-1}(\psi))\,|\di g^{-1}(\psi)/\di\psi|$. The coordinates of each model are listed in Sections~\ref{ss:design}, \ref{ss:mcfx} and \ref{ss:app2models}, and $\log p_{\bm{\psi}}$ is the sum over the coordinates of the terms below. Six cases cover all the models.

A parameter with a normal prior enters as itself and adds no Jacobian: the levels $\mu$, $\mu_1$ and $\mu_2$, the coefficients $\lambda$ and $\phi_{21}$, the time-of-day coefficients $\kappa_1, \ldots, \kappa_4$, the mean $\mu_\xi$ of the volatility jump, and the degrees of freedom through $\psi = \log(\nu - 2)$, on which the prior $\distn{N}(2, 1)$ is placed directly. Under this prior $\nu - 2$ is log-normal, with density
\[
	p(\nu) = \frac{1}{(\nu - 2)\sqrt{2\pi}}\exp\left\{-\frac{(\log(\nu - 2) - 2)^2}{2}\right\}, \qquad \nu > 2,
\]
whose median is $2 + \e^{2} \approx 9.4$ and whose mean is $2 + \e^{5/2} \approx 14.2$.

A persistence parameter with prior $(1 + \phi)/2 \sim \distn{B}(a, b)$ enters as $z = \text{atanh}\,\phi$. Since $1 + \tanh z = \e^{z}/\cosh z$, $1 - \tanh z = \e^{-z}/\cosh z$ and $\di\phi/\di z = \cosh^{-2} z$,
\[
	\log p(z) = (a - b)z - (a + b)\log\cosh z - (a + b - 1)\log 2 - \log B(a, b),
\]
where $B(a, b)$ is the beta function. This case covers $\phi$, $\phi_1$ and $\phi_2$ in the models with independent persistence parameters.

A variance with prior $\sigma^2 \sim \distn{IG}(\nu_0, S_0)$, whose density is $S_0^{\nu_0}\Gamma(\nu_0)^{-1}(\sigma^2)^{-\nu_0 - 1}\e^{-S_0/\sigma^2}$, enters as $s = \log\sigma^2$. Since $\di\sigma^2/\di s = \e^{s}$,
\[
	\log p(s) = \nu_0\log S_0 - \log\Gamma(\nu_0) - \nu_0 s - S_0\e^{-s}.
\]
This case covers $\sigma^2$, $\sigma_1^2$, $\sigma_2^2$, $\sigma_\zeta^2$ and $\sigma_\xi^2$.

The correlation with prior $\rho \sim \distn{U}(-1, 1)$ enters as $r = \text{atanh}\,\rho$, with $\di\rho/\di r = \cosh^{-2} r$, so $\log p(r) = -\log 2 - 2\log\cosh r$.

The jump probability with prior $\omega \sim \distn{B}(a_\omega, b_\omega)$ enters as $k = \text{logit}\,\omega$. Since $\di\omega/\di k = \omega(1 - \omega)$, $\log\omega = k - \log(1 + \e^{k})$ and $\log(1 - \omega) = -\log(1 + \e^{k})$,
\[
	\log p(k) = a_\omega k - (a_\omega + b_\omega)\log(1 + \e^{k}) - \log B(a_\omega, b_\omega).
\]

The second persistence parameter of the two-factor models, with prior $\phi_2 \gvn \phi_1 \sim \distn{U}(-1, \phi_1)$ and density $1/(1 + \phi_1)$, enters as $w = \log(z_1 - z_2)$ with $z_i = \text{atanh}\,\phi_i$, so that $\phi_2 = \tanh(z_1 - \e^{w})$ and $\partial\phi_2/\partial w = -\e^{w}\cosh^{-2}(z_1 - \e^{w})$. The conditional prior density of $w$ given $z_1$ is
\[
	\log p(w \gvn z_1) = w - 2\log\cosh(z_1 - \e^{w}) - \log(1 + \tanh z_1),
\]
and $z_1$ itself carries the term of the persistence case above.

\section{Monthly Bayes Factors}\label{app:diagnostic}
This appendix reports the monthly estimates behind Section~\ref{ss:app2recursive} and the importance-sampling estimates at the three month-end fits at which $|\Delta_2|$ exceeds one. Both models with such fits have $d = 11$ parameters.

\subsection{Monthly Predictive Comparisons}
Table~\ref{tab:recursive} gives the log Bayes factors against SV by period. On the training sample, every model but SVJ is already favored over SV by 10 to 12, and the ranking over the ten months from March to December is close to the full-sample one. The last column shows that $|\Delta_2|$ does not exceed 0.45 at any month-end fit but three: SVJV at the end of June, with $\Delta_2 = 3.41$ at $t = 9{,}516$, and TFJ at the end of August and of September, with $1.49$ and $-3.27$ at $t = 12{,}792$ and $14{,}430$. 

At these three fits the table uses the importance-sampling estimate of Section~\ref{app:is} in place of $\log\hat m_2$. The replacement lowers the June entry of SVJV from 7.8 to 4.3 and raises its July entry from $-0.1$ to 3.3; for TFJ it moves about 1.4 from August to September and about 3.0 from October to September. A month-end value enters the increment of its own month with a positive sign and that of the following month with a negative sign, so the cumulative row does not depend on the three values.

% The five marked entries are derived from the rounded differences between the importance-sampling and the repaired estimates (0.09, 0.02, -0.01); regenerate them from the unrounded results.
\begin{table}[H]
	\caption{Log Bayes factors against SV by period, five-minute SPY returns of 2025. The first row is the training sample, each later row is the predictive log Bayes factor of the month, and the last two rows are the sum over March to December and the full-sample value. The last column is the largest $|\Delta_2|$ among the seven fits at the month end. Entries marked $^{a}$ use the importance-sampling estimate of Section~\ref{app:is}.}
	\label{tab:recursive}
	\centering
	\footnotesize\setlength{\tabcolsep}{5pt}
	\begin{tabular}{lrrrrrrc}
		\toprule
		Period & SV-$t$ & SVJ & SVJV & TF & TF-$t$ & TFJ & $\max|\Delta_2|$ \\
		\midrule
		Training, to Feb 28 & 10.39 & 5.44 & 11.23 & 12.23 & 11.93 & 10.29 & 0.32 \\
		March     & $-2.30$ & 1.70 & 0.53 & $-1.43$ & $-2.36$ & $-0.13$ & 0.37 \\
		April     & 0.84 & 0.05 & 8.71 & 1.90 & 1.89 & 2.28 & 0.35 \\
		May       & 7.17 & 4.95 & 11.51 & 9.62 & 10.10 & 9.54 & 0.45 \\
		June      & 4.20 & 3.38 & 4.31$^{a}$ & 8.11 & 7.78 & 8.40 & 3.41 \\
		July      & 1.53 & $-4.03$ & 3.34$^{a}$ & 10.02 & 9.19 & 7.58 & 0.41 \\
		August    & 8.90 & 2.55 & 5.93 & 14.11 & 14.45 & 13.51$^{a}$ & 1.49 \\
		September & $-4.63$ & $-5.41$ & 3.29 & 2.80 & 1.88 & 1.59$^{a}$ & 3.27 \\
		October   & 1.14 & 5.35 & 20.80 & 7.17 & 6.95 & 10.73$^{a}$ & 0.23 \\
		November, to Nov 26 & $-2.06$ & $-0.40$ & $-1.19$ & 3.67 & 2.72 & 3.50 & 0.26 \\
		December, from Nov 27 & $-0.09$ & $-1.33$ & 0.41 & 4.97 & 4.86 & 4.49 & 0.31 \\
		\midrule
		March to December & 14.70 & 6.82 & 57.65 & 60.93 & 57.49 & 61.48 & \\
		Full sample & 25.08 & 12.25 & 68.88 & 73.16 & 69.42 & 71.77 & \\
		\bottomrule
	\end{tabular}
\end{table}

\subsection{The Three Fits with a Large Second-Order Term} \label{app:is}

The large second-order terms at the three fits are not numerical artifacts: a second implementation of the grid filter reproduces $\Delta_2$ to $10^{-4}$ at the two TFJ fits. Recomputing \eqref{eq:delta2} without one coordinate at a time attributes almost all of $\Delta_2$ to the log variance of the jump sizes, $\log\sigma_\xi^2$ in SVJV and $\log\sigma_\zeta^2$ in TFJ (see Table~\ref{tab:flagged}).

At the three fits we estimate $\Em[\e^{R(\bG)}]$ in \eqref{eq:ratio} by importance sampling over the static parameters, as described in Section~\ref{ss:budget}. The proposal is a multivariate Student-$t$ distribution with five degrees of freedom in the standardized coordinates, and we use 2{,}000 draws.\footnote{A draw at which a parameter falls outside the range of the grid receives zero weight; this happens at 27 of the 2{,}000 draws at the SVJV fit of June and at most 10 draws elsewhere.} With draws $\bu^{(m)}$ from the proposal density $q$, the average of $\e^{R(\bu^{(m)})}\varphi(\bu^{(m)})/q(\bu^{(m)})$ estimates $m_T/m_1$, where $\varphi$ is the standard normal density.

We highlight two results. First, the importance sampler is accurate: the numerical standard error is 0.04 at each of the three fits, and at four adjacent month-end fits of the same models, where $|\Delta_2|$ is below one, the importance-sampling estimate is within 0.15 of $\log\hat m_2$. Second, at the three fits the estimate differs from $\log\hat m_2$ by about $-\Delta_2$ (see Table~\ref{tab:flagged}), so the second-order term correctly identifies the fits at which the expansion should not be used.

\begin{table}[H]
	\caption{The three month-end fits at which $|\Delta_2|$ exceeds one. Omission is the change in $\Delta_2$ when the log jump variance is left out of \eqref{eq:delta2}, Std.\ dev.\ is its local Gaussian standard deviation, and IS is the importance-sampling estimate of $\log m_T$ (Student-$t$ proposal with five degrees of freedom, 2{,}000 draws, numerical standard error 0.04).}
	\label{tab:flagged}
	\centering
	\footnotesize\setlength{\tabcolsep}{5pt}
	\begin{tabular}{llrrrrrr}
		\toprule
		Model & Date & $t$ & $\Delta_2$ & Omission & Std.\ dev. & IS $-\log\hat m_1$ & IS $-\log\hat m_2$ \\
		\midrule
		SVJV & June 30 & 9{,}516 & 3.41 & 3.28 & 1.23 & $-0.06$ & $-3.47$ \\
		TFJ & August 29 & 12{,}792 & 1.49 & 1.36 & 0.79 & 0.05 & $-1.43$ \\
		TFJ & September 30 & 14{,}430 & $-3.27$ & $-3.38$ & 1.12 & $-0.31$ & 2.97 \\
		\bottomrule
	\end{tabular}
\end{table}
\end{document}